\documentclass[a4paper,fleqn,usenatbib]{mnras}

\usepackage{lineno}
\usepackage[T1]{fontenc}
\usepackage{ae,aecompl}
\usepackage{textcomp}
\usepackage{graphicx}	
\usepackage{amsmath}	
\usepackage{amssymb}	
\usepackage{dirtytalk}
\usepackage{siunitx}
\usepackage{tablefootnote}

\usepackage{xcolor}

\title[Progenitor mass constraints for two type Ib SNe]{Progenitor mass constraints for the type Ib intermediate-luminosity SN~2015ap and the highly extinguished SN~2016bau}

\author[Aryan et al.]
{Amar Aryan\thanks{e-mail :amar@aries.res.in,amararyan941@gmail.com}$^{1,2}$,
S. B. Pandey$^{1}$,
WeiKang Zheng$^{3}$,
Alexei V. Filippenko$^{3,4}$,
\newauthor
 Jozsef Vinko$^{5,6,7,8}$, Ryoma Ouchi$^{9}$,
Isaac Shivvers$^{3}$, Heechan Yuk$^{3,10}$, Sahana Kumar$^{3,11}$,
\newauthor
  Samantha Stegman$^{3}$, Goni Halevi$^{3,12}$, Timothy W. Ross$^{3}$, Carolina Gould$^{3}$,
\newauthor
Sameen Yunus$^{3}$, Raphael Baer-Way$^{3}$, Asia deGraw$^{3}$, Keiichi Maeda$^{9}$,  
\newauthor
D. Bhattacharya$^{13}$, Amit Kumar$^{1,14}$, Rahul Gupta$^{1,2}$, Abhay P. Yadav$^{15}$, 
\newauthor
David A. H. Buckley$^{16}$, Kuntal Misra$^{1}$, and S. N. Tiwari$^{2}$
\\
$^{1}$Aryabhatta Research Institute of Observational Sciences, Manora Peak, Nainital 263 002 India\\
$^{2}$Department of Physics, Deen Dayal Upadhyay Gorakhpur University, Gorakhpur, Civil Lines, Gorakhpur (U.P.)  273 009, India\\
$^{3}$Department of Astronomy, University of California, Berkeley, CA 94720-3411, USA\\
$^{4}$Miller Institute for Basic Research in Science, University of California,
Berkeley, CA 94720, USA\\
$^{5}$Department of Astronomy, University of Texas at Austin, Austin, TX, USA \\
$^{6}$CSFK Konkoly Observatory, Konkoly Thege M. ut 15-17, Budapest, 1121, Hungary\\
$^{7}$Department of Optics and Quantum Electronics, University of Szeged, D\'om t\'er 9, Szeged, 6720 Hungary\\
$^{8}$ELTE E\"otv\"os Lor\'and University, Institute of Physics, P\'azm\'any P\'eter s\'et\'any 1/A, Budapest, 1117 Hungary\\
$^{9}$Department of Astronomy, Kyoto University, Kitashirakawa-Oiwake-cho, Sakyo-ku, Kyoto 606-8502, Japan\\
$^{10}$Department of Physics and Astronomy, University of Oklahoma, 440 W.
Brooks St., Norman, OK 73019, USA\\
$^{11}$Department of Physics, Florida State University, Tallahassee, FL 32306,
USA\\
$^{12}$Department of Astrophysical Sciences, Princeton University, 4 Ivy Lane,
Princeton, NJ 08540, USA\\
$^{13}$The Inter-University Centre for Astronomy and Astrophysics (IUCAA), Ganeshkhind, Savitribai Phule Pune\\ University Campus, Pune 411 007, India\\
$^{14}$School of Studies in Physics and Astrophysics, Pt. Ravishankar Shukla University, Chattisgarh 492 010, India\\
$^{15}$Department of Physics and Astronomy, National Institute of Technology Rourkela-769008, Odisha, India\\
$^{16}$South African Astronomical Observatory, PO Box 9, Observatory 7935, Cape Town, South Africa
}

\date{Accepted 2021 May 11. Received 2021 May 11; in original form 2020 December 24}

\pubyear{2021}

\begin{document}
\label{firstpage}
\pagerange{\pageref{firstpage}--\pageref{lastpage}}
\maketitle

\begin{abstract}
Photometric and spectroscopic analyses of the intermediate-luminosity Type Ib supernova (SN) 2015ap and of the heavily reddened Type Ib SN~2016bau are discussed. Photometric properties of the two SNe, such as colour evolution, bolometric luminosity, photospheric radius, temperature, and velocity evolution, are also constrained. The ejecta mass, synthesised nickel mass, and  kinetic energy of the ejecta are calculated from their light-curve analysis. We also model and compare the spectra of SN~2015ap and SN~2016bau at various stages of their evolution. The P~Cygni profiles of various lines present in the spectra are used to determine the velocity evolution of the ejecta. To account for the observed photometric and spectroscopic properties of the two SNe, we have computed 12\,$M_\odot$ zero-age main sequence (ZAMS) star models and evolved them  until the onset of core collapse using the publicly available stellar-evolution code {\tt MESA}. Synthetic explosions were produced using the public version of  {\tt STELLA} and another publicly available code, {\tt SNEC}, utilising the {\tt MESA} models. {\tt SNEC} and {\tt STELLA} provide various observable properties such as the bolometric luminosity and velocity evolution. The parameters produced by {\tt SNEC}/{\tt STELLA} and our observations show close agreement with each other, thus supporting a  12\,$M_\odot$ ZAMS star as the possible progenitor for SN~2015ap, while the progenitor of SN~2016bau is slightly less massive, being close to the boundary between SN and non-SN as the final product.
\end{abstract}


\section{Introduction}
\label{sec:Introduction}
Core-collapse supernovae (CCSNe) are among the most powerful  astronomical explosions, occurring during the final stellar evolutionary stages of massive stars ($M > 8$--10\,$M_\odot$;  e.g., \citealt[][]{Garry2004, Woosley2005, Groh2017}). 
Extensive reviews of various types of CCSNe and criteria used to categorise them are provided by (among others) \citet[][]{Filippenko1997} and \citet[][]{Gal-Yam16}.
CCSNe are broadly classified according to the presence or absence of hydrogen (H) features in their spectra. SNe showing prominent H features are classified as Type II, while those lacking them are  Type I. These classes are further divided into various subclasses. Type Ib SNe exhibit prominent helium (He) features in their spectra, whereas Type Ic SNe show neither H nor He obvious features. Prominent features of intermediate-mass elements such as O, Mg, and Ca are also seen in Type Ib and Type Ic SN spectra. Although SNe~Ib/c lack obvious H features in their  early-time spectra, few studies have focused on the existence of H in SNe~Ib \citep[e.g.,][]{Branch2002, Branch2006, Hachinger2012, Elmhamdi2006}. Type IIb SNe form a transition class of objects that link SNe~II and SNe~Ib \citep[][]{Filippenko1988, Filippenko1993, Smartt2009}. The early-phase spectra of SNe~IIb display prominent H features, while unambiguous He features appear after a few weeks.

The main powering mechanism in normal SNe~Ib/c is radioactive decay of $^{56}$Ni and $^{56}$Co, leading to the deposition of energetic gamma rays that thermalise in the homologously expanding ejecta \citep[e.g.,][]{Arnett1980, Arnett1982, Arnett1996, Nadyozhin1994, Chatzopoulous2013, Nicholl2017}. Some SNe~Ib (e.g., SN2005bf, \citealt[][]{Maeda2007}) have also shown evidence for the light curves being powered by the spin-down of a young magnetar  \citep[e.g.,][]{Ostriker1971, Arnett1989, Maeda2007, Kasen2010, Woosley2010, Chatzopoulous2013, Nicholl2017}. In the post-photospheric phase, when the SN ejecta become optically thin, the light curves of CCSNe are powered by energy deposition from the radioactive decay of $^{56}$Ni to $^{56}$Co  and finally to $^{56}$Fe. In many cases, however, the SN progenitor is embedded within dense circumstellar matter (CSM), so when the SN explosion occurs the SN ejecta may violently interact with the CSM, resulting in the formation of forward and reverse shocks that deposit their kinetic energy into the material which is radiatively released and thus powers the light curve \citep[e.g.,][]{ Chevalier1982, Chevalier1994, Moriya2011, Ginzberg2012, Chatzopoulous2013, Nicholl2017}.

Understanding the possible progenitors of H-stripped CCSNe is still a challenging task, though a few studies have been performed. For SNe~Ib/c, broadly two scenarios are proposed. The first involves relatively low-mass progenitors ($> 11\,M_\odot$) in binary systems  \citep[][]{Podsiadlowski1992, Nomoto1995, Smartt2009}, where the primary star lost its H envelope through transfer of mass to a companion star. The second considers massive Wolf-Rayet (WR) stars ($> 20$--25\, $M_\odot$) that lose mass via stellar winds \citep[e.g.,][]{Gaskell1986, Eldridge2011, Groh2013}. In one relatively recent study, \citet[][]{Cao2013} reported a possible progenitor of SN iPTF13bvn identified in pre-explosion images within a 2$\sigma$ error radius, consistent with a massive WR progenitor star. The massive WR progenitor scenario is also supported by stellar evolutionary models. Based on observational evidence, from early- and nebular-phase spectroscopy of  SNe~Ib, both massive WR stars as well as interacting binary progenitors are proposed. For SNe~IIb, the direct detections of objects in pre-explosion images of four cases also indicate either massive WR stars ($M_{\rm ZAMS} \approx 10$--28\,$M_\odot$; \citealt{Crockett2008}) or more extended yellow supergiants (YSGs) with $M_{\rm ZAMS} = 12$--17\,$M_\odot$ \citep[][]{Van2013, Folatelli2014, Smartt2015} as possible progenitors. The hydrodynamical modelling of the possible progenitors (identified either via direct imaging as in the case of iptf13bvn \citep[][]{Cao2013} or indirect methods which include nebular-phase spectral modelling \citep[][]{Jerkstrand2015, Uomoto1986}) and simulating their synthetic explosions can be vital to understanding their nature, physical conditions, circumstellar environment, and chemical compositions. Unfortunately, only a handful of such studies have been performed in the cases of stripped-envelope SNe, including the Type Ib SN iptf13bvn \citep[][]{Cao2013, Bersten2014, Paxton2018}, the famous Type IIb SN~2016gkg \citep[][]{Bersten2018}, and the Type IIb SN~2011dh \citep[][]{Bersten2012}. Our work takes such studies one step further as we perform hydrodynamical simulations of the possible progenitors of two SNe~Ib and also simulate their synthetic explosions.

In this paper, we explore the photometric and spectroscopic behaviour of SN~2015ap and SN~2016bau, primarily using  data obtained using the KAIT, Nickel, and Shane telescopes at Lick Observatory. Based on the analysis, we model and attempt to place constraints on the properties of the progenitors of these two SNe. In Sec.~\ref{sec:Data_red}, details about various telescopes and reduction procedures are presented. Sec.~\ref{sec:Photometric} provides methods to correct for the Milky Way and the host-galaxy extinction. Photometric properties of the two SNe, such as their bolometric light curve, temperature, radius, and velocity evolution, are also discussed. Sec.~\ref{sec:Spectral} includes the analysis describing the spectral evolution of SN~2015ap and SN~2016bau, as well as comparisons with other similar and well-studied SNe; we also model the spectra of these SNe using {\tt SYN++}. Quasi-bolometric light-curve modelling is performed and discussed in Sec.~\ref{sec:lc_model}; light curves corresponding to various powering mechanisms of SNe are fitted to the observed quasi-bolometric light curves. The assumptions and methods for modelling the possible progenitors of the two SNe and their evolution until the onset of core-collapse using MESA are presented in  Sec.~\ref{sec:mesa}. We discuss the assumptions and methods for producing the synthetic explosions using SNEC and STELLA in Sec.~\ref{sec:snec}; here, the comparisons between the parameters obtained through synthetic explosions and observed ones are presented. In Sec.~\ref{sec:Discussions}, we discuss the major results and findings of our analysis. Finally, in Sec.~\ref{sec:Conclusions}, we summarise our work by briefly discussing the major outcomes and provide concluding remarks. 

\begin{figure*}
\centering
\includegraphics[width=\textwidth]{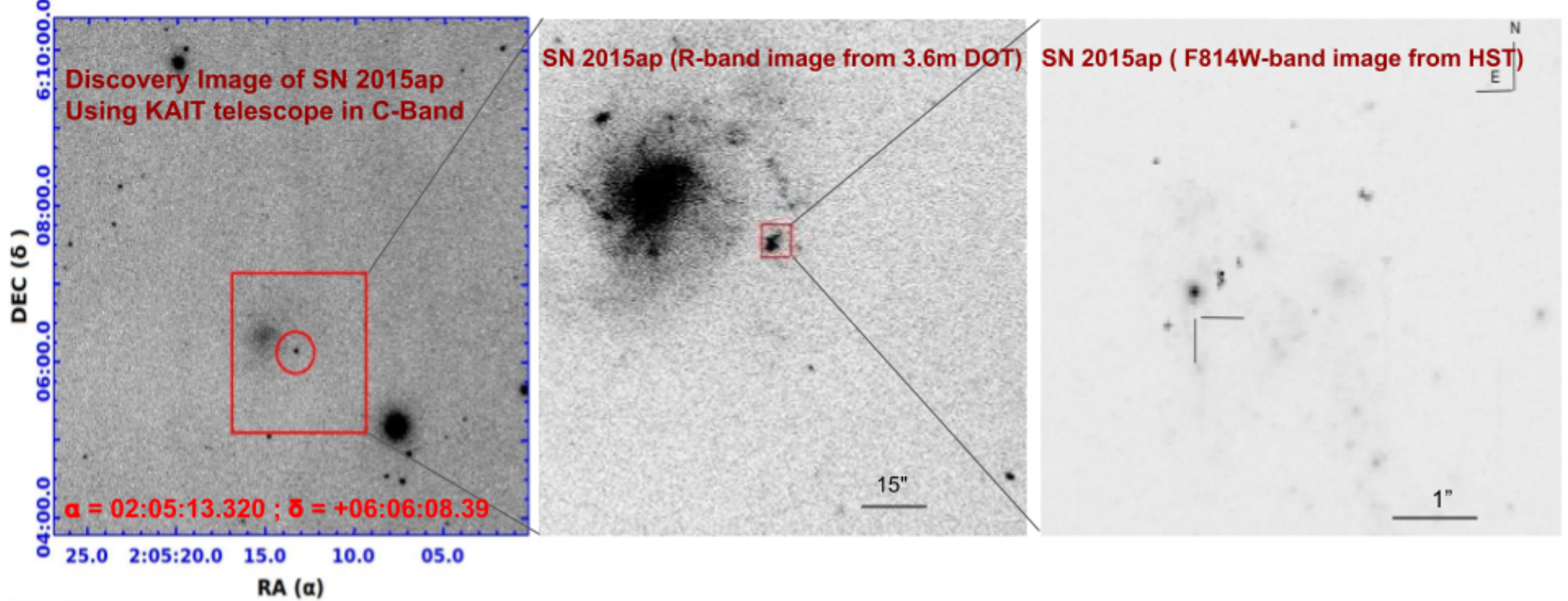}
\caption{Left panel: The discovery image of SN~2015ap using the KAIT telescope in C-filter.The SN is marked inside the red circle. . Middle panel: A zoomed version of an $R$-band image with $2' \times 2'$ FoV obtained on 6 Oct. 2020 using a 4k $\times$ 4k CCD imager mounted on the 3.6\,m DOT. Right panel: A further zoomed, about $6'' \times 6''$ FoV of the location of SN~2015ap taken on 12 Aug. 2020 using the {\it Hubble Space Telescope (HST)}  in the F814W filter. The SN seems to be fainter than the detection limit of {\it HST} at this time. A log of {\it HST} observations is included in the Appendix.}
\label{fig:finding_chart15ap}
\end{figure*}

\section{Data acquisition and reduction }
\label{sec:Data_red}

SN~2015ap was discovered by KAIT as part of the Lick Observatory Supernova Search \citep[LOSS;][]{Filippenko2001}, in an 18\,s unfiltered image (close to the $R$ band; see \citealt{Li2003}) taken at  11:16:31 on 2015~Sep.~08 \citep[][]{Ross2015}, at $17.41\pm0.06$\,mag. The object was also marginally detected one day earlier on Sep. 07.47 with $18.07\pm0.22$\, mag. We measure its J2000.0 coordinates to be $\alpha=02^{\mathrm{h}}05^{\mathrm{m}}13.32^{\mathrm{s}}$, $\delta=+06^{\circ}06\arcmin08\farcs4$, with an uncertainty of $0\farcs5$ in each coordinate. SN~2015ap is $28\farcs3$ west and $16\farcs2$ south of the nucleus of its host galaxy IC~1776, which has a redshift of $z=0.011375 \pm 0.000017$ \citep[][]{Chengalur1993} and a barred-spiral morphology \citep[SB(s)d;][]{deVaucouleurs1991}). SN~2016bau was discovered by Ron Arbour in an unfiltered image taken at 23:22:33 on 2016~Mar.~13  \citep[][]{2016TNSTR.215....1A}, at 17.8\,mag.  Its J2000.0 coordinates are given as $\alpha=11^{\mathrm{h}}20^{\mathrm{m}}59.02{\mathrm{s}}$, $\delta=+53^{\circ}10\arcmin25\farcs6$. SN~2016bau is $35\farcs3$ west and $15\farcs2$ north of the nucleus of its host galaxy NGC~3631, which has $z=0.00384 \pm 0.00014$ \citep[][]{Falco1999} and a  morphology of SAc~C \citep[][]{Ann2015}. Both of these SNe were classified as SN~Ib on the basis of well-developed features of He~I, Fe~II (blended), and Ca~II, a few days after maximum brightness.

Figure~\ref{fig:finding_chart15ap} shows finder charts of SN~2015ap obtained with different telescopes. 
The $B$, $V$, $R$, and $I$ follow-up images of SN~2015ap and SN~2016bau were obtained with both the 0.76\,m Katzman Automatic Imaging Telescope \citep[KAIT;][]{Filippenko2001} and the 1\,m Nickel telescope at Lick Observatory. All images were reduced using a custom pipeline\footnote{https://github.com/benstahl92/LOSSPhotPypeline} detailed by \citet[][]{Stahl2019}. Here, we briefly summarise the process for photometry. The image-subtraction procedures were applied in order to remove the host-galaxy light, using additional images obtained after the SN had faded below our detection limit. Point-spread-function (PSF) photometry was obtained using DAOPHOT \citep[][]{Stetson1987} from the IDL Astronomy User\textquotesingle s Library\footnote{http://idlastro.gsfc.nasa.gov/}.
Three nearby stars were chosen from the Pan-STARRS1\footnote{http://archive.stsci.edu/panstarrs/search.php} catalogue for calibration. Their magnitudes were first transformed into \citet{Landolt1992} magnitudes using the empirical prescription presented by \citet[][Eq. 6]{Torny2012} and then transformed to the KAIT/Nickel natural system. Apparent magnitudes were all measured in the KAIT4/Nickel2 natural system. The final results were transformed to the standard system using local calibrators and colour terms for KAIT4 and Nickel2 \citep[][]{Stahl2019}. 
In addition, the $U$-band photometry was obtained from the {\it Swift} Optical/Ultraviolet Supernova Archive (SOUSA; https://archive.stsci.edu/prepds/sousa/; \citealt[][]{Brown2014}).

There are a total of 17 spectra of SN~2015ap; 15 of them were taken with the Kast spectrograph\footnote{https://mthamilton.ucolick.org/techdocs/instruments/kast} \citep{Miller&Stone1993} on the 3\,m Shane telescope at Lick Observatory and the other 2 were taken with LRIS\footnote{https://www2.keck.hawaii.edu/inst/lris/lrishome.html} \citep{Oke1995} on the Keck-I 10\,m telescope. Except for the earliest spectrum taken on 10.531 Sep. 2015 (UT dates are used throughout this paper) using the Lick Shane/Kast system, 16 spectra in this paper were published by \citet[][]{Shivvers2018}; refer to that paper for details of the observations. Also, there are 8 spectra of SN~2016bau, all obtained with the Lick Shane/Kast system. We used  the 600/4310 grism on the blue side, the 300/7500 grating on the red side, the d5500 dichroic, and a long slit $2''$ wide. For LRIS we used the 600/4000 grism on the blue side, the 400/8500 grating on the red side, and a $1''$ slit. The spectra were binned to  2\,\si{\angstrom}\ pixel$^{-1}$.

\section{Photometric Properties}
\label{sec:Photometric}
In this section, we discuss various photometric properties of SN~2015ap and SN~2016bau, including their colour evolution, extinction, quasi-bolometric light curves, and various blackbody parameters.

\begin{figure}
	\includegraphics[width=\columnwidth]{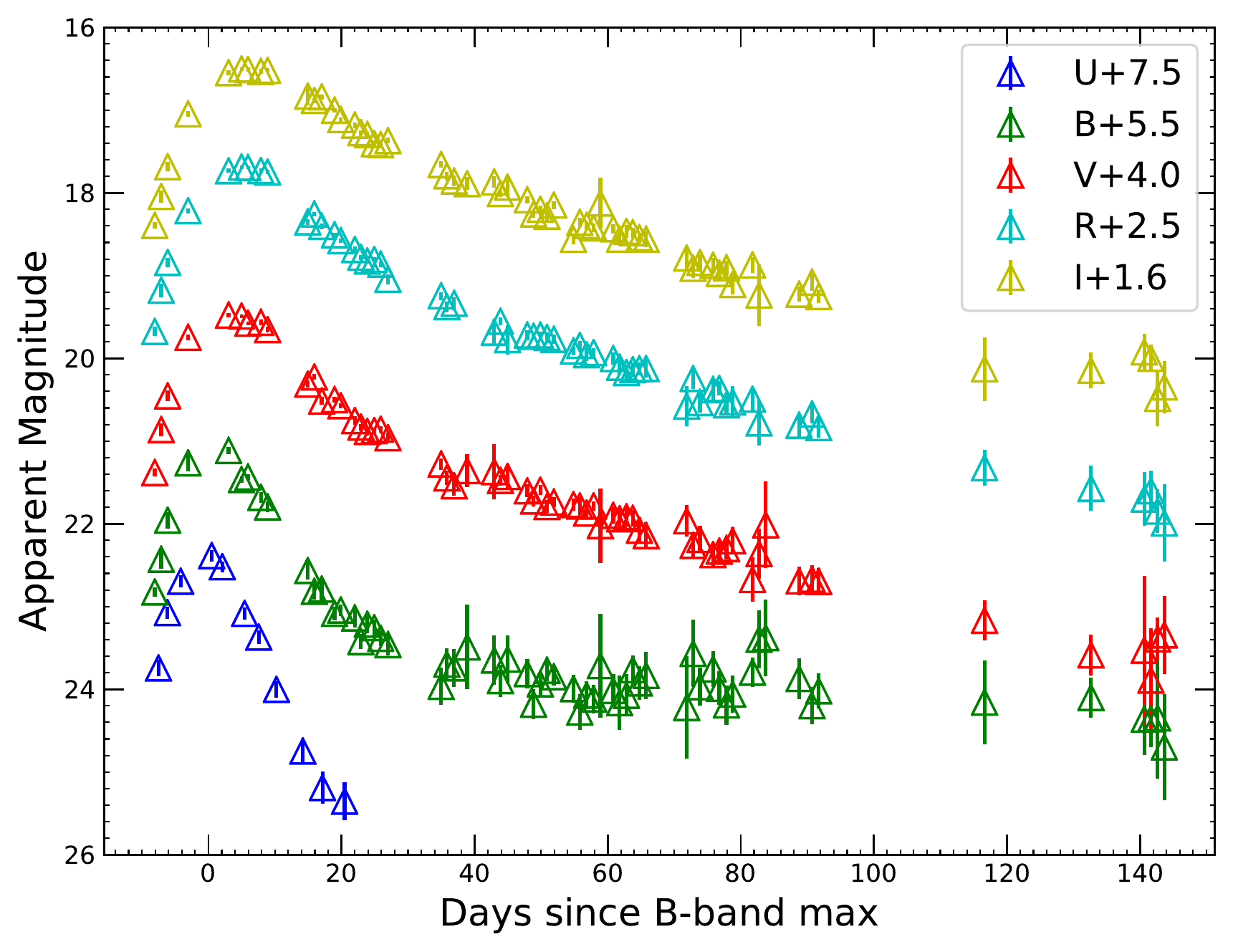}
    \caption{$UBVRI$ light curves of SN~2015ap, where $BVRI$ data were obtained with KAIT while the $U$-band data were taken from the UVOT mounted on {\it Swift} (https://swift.gsfc.nasa.gov).}
    \label{fig:UBVRI_LC}
\end{figure}

\begin{figure}	
    \includegraphics[width=\columnwidth]{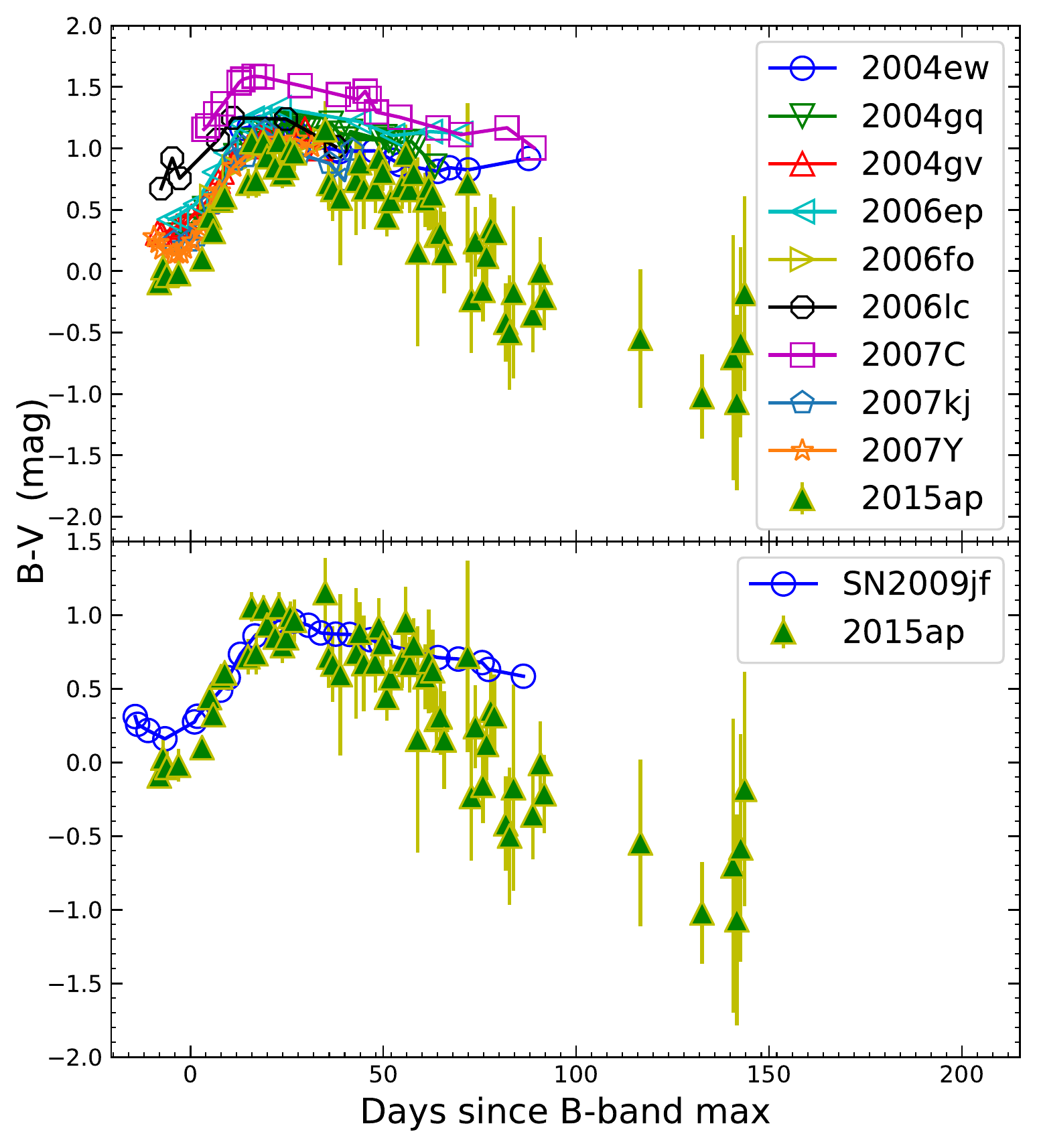}
    \caption{The top panel shows a comparison of the $(B-V)$ colour of SN~2015ap (corrected for Milky Way extinction) with that of other Type Ib SNe. The data for SNe other than SN~2015ap are taken from \citet[][]{Stritzinger2018}. The bottom panel shows the $(B-V)$ colour curves of SN~2015ap and SN~2009jf (both corrected for Milky Way extinction).}
    \label{fig:color_curve_2015ap}
\end{figure}

\subsection{Photometric properties of SN2015ap}
\label{subsec:Photometric_SN2015ap}

Most of the analysis in this paper has been performed with respect to $B$-band maximum brightness. To find its date, we fit a sixth-order polynomial to the $B$ data which well sample the photospheric phase. The resulting date of $B$-band maximum is MJD $57282.47\pm2.56$. 
To determine the explosion epoch ($t_{\rm exp}$) of SN~2015ap, we use $R$-band data. The $R$ magnitudes are converted into fluxes and a sixth-order polynomial is fitted. We extrapolate the polynomial and the epoch corresponding to zero flux is taken as the explosion epoch, MJD $57272.72 \pm 1.49$, which is in good agreement with \citet[][]{Prentice2019}.

Figure~\ref{fig:UBVRI_LC} shows the $UBVRI$ light curves of SN~2015ap. The rise rate for the $U$-band light curve is faster than that of other bands, and similarly the $U$ decline rate is faster, making the $U$ light curve much narrower compared to other bands. As we go to longer wavelengths, we see that the light curves become broader, with the $I$-band light curve being the broadest.

\subsubsection{Colour evolution and extinction correction}

Distances are taken from the NASA Extragalctic Database\footnote{https://ned.ipac.caltech.edu/} (NED), and a cosmological model with H$_{0} = 73.8$\,km\,s$^{-1}$\,Mpc$^{-1}$, $\Omega_{m}=0.3$, and $\Omega_{\Lambda} = 0.7$ is assumed throughout.
For SN~2015ap, we corrected for Milky Way (MW) extinction using NED following \citet[][]{Schlafly2011}. In the direction of SN~2015ap, the Galactic extinctions for the $U$, $B$, $V$, $R$, and $I$ bands are 0.185, 0.154, 0.117, 0.092, and 0.064\,mag, respectively. 

The top panel of Figure~\ref{fig:color_curve_2015ap} shows a comparison of the $(B-V)$ colour of SN~2015ap with that of other Type Ib SNe (all corrected for MW extinction). SN~2015ap seems to be the least reddened and lies below nearly all of the other SNe~Ib.

Following \citet[][]{Prentice2019}, the host-galaxy contamination is negligible and hence ignored. To further support this assumption, the $(B-V)$ colour curve of SN~2009jf \citep[][]{Sahu2011} is shown in the bottom panel of Figure~\ref{fig:color_curve_2015ap}, corrected for an MW colour excess of $E(B-V)_{\rm MW} = 0.112$\,mag (host extinction is negligible). In order to match the ($B-V$) colour curve of SN~2009jf, we need not apply any shift to the MW-corrected SN~2015ap $(B-V)$ colour curve.

\begin{figure}
	\includegraphics[width=\columnwidth]{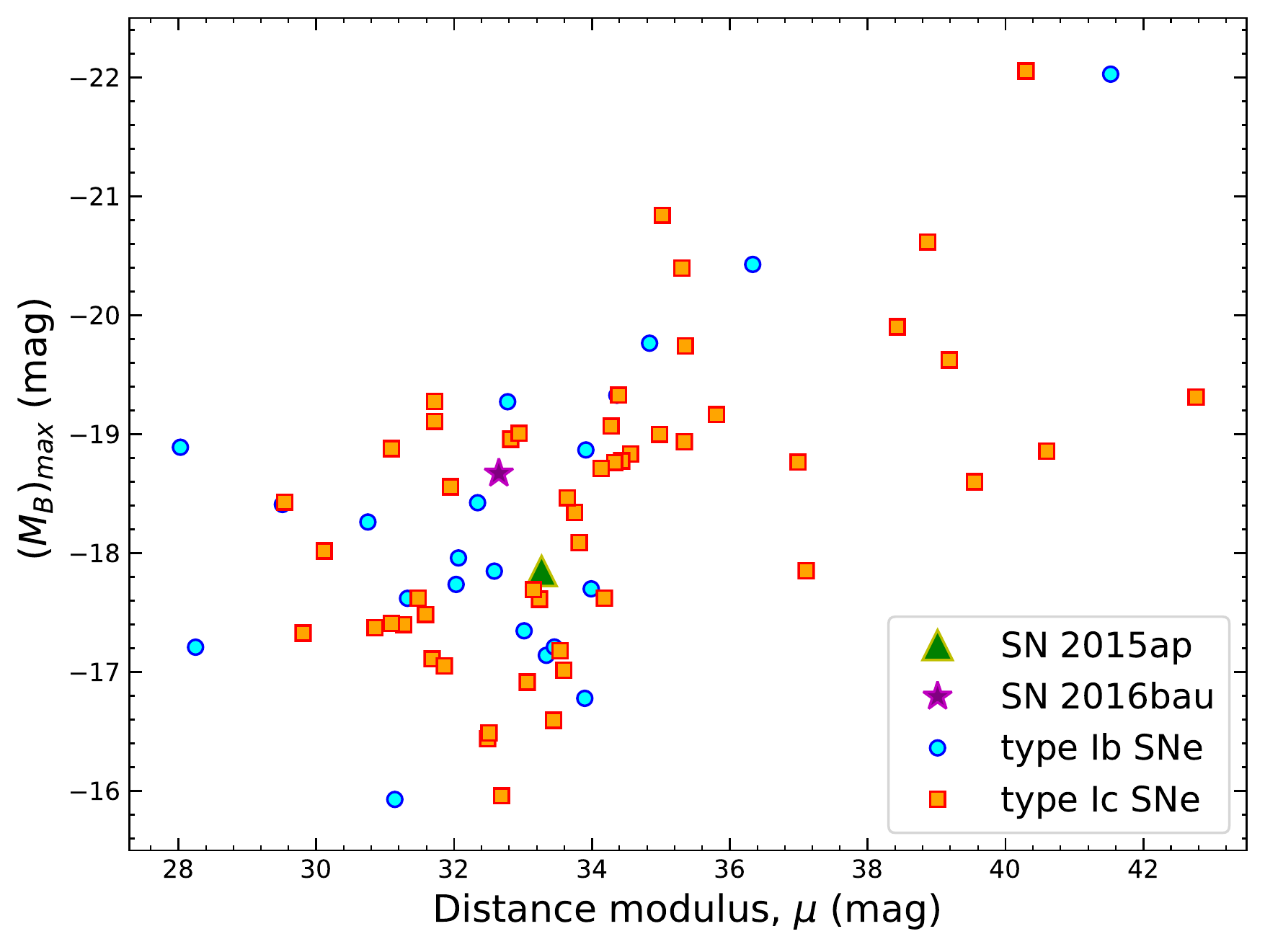}
    \caption{The position of SN~2015ap and SN~2016bau in the Miller diagram.}
    \label{fig:miller}
\end{figure}

Figure~\ref{fig:miller} illustrates the position of SN~2015ap in the Miller diagram \citep[][]{Richardson2014}. The distance modulus ($\mu$) of SN~2015ap is 33.269\,mag; thus, SN~2015ap appears to be a normal SN~Ib. For SN~2016bau, we calculate $\mu = 32.65$\,mag; based on its position in Figure~\ref{fig:miller}, it seems to be a moderately luminous, normal SN~Ib.

\subsubsection{Quasi-bolometric and bolometric light curves}

To obtain the quasi-bolometric light curves, we made use of the {\tt superbol} code \citep{Nicholl2018}. We first provided the extinction-corrected $U$, $B$, $V$, $R$, and $I$ data to {\tt superbol}. Thereafter, it mapped the light curve in each filter to a common set of times through the processes of interpolation and extrapolation. It then fit blackbodies to the spectral energy distribution (SED) at each epoch, up to the observed wavelength, to give the quasi-bolometric light curve by performing trapezoidal integration. The peak quasi-bolometric luminosity obtained through integrating the flux over a wavelength range of 4000--10,000\,\si{\angstrom}\ is $10^{(42.548\pm0.019)}$\,erg\,s$^{-1}$, in agreement with \citet[][]{Prentice2019}.

\begin{figure}
	\includegraphics[width=\columnwidth]{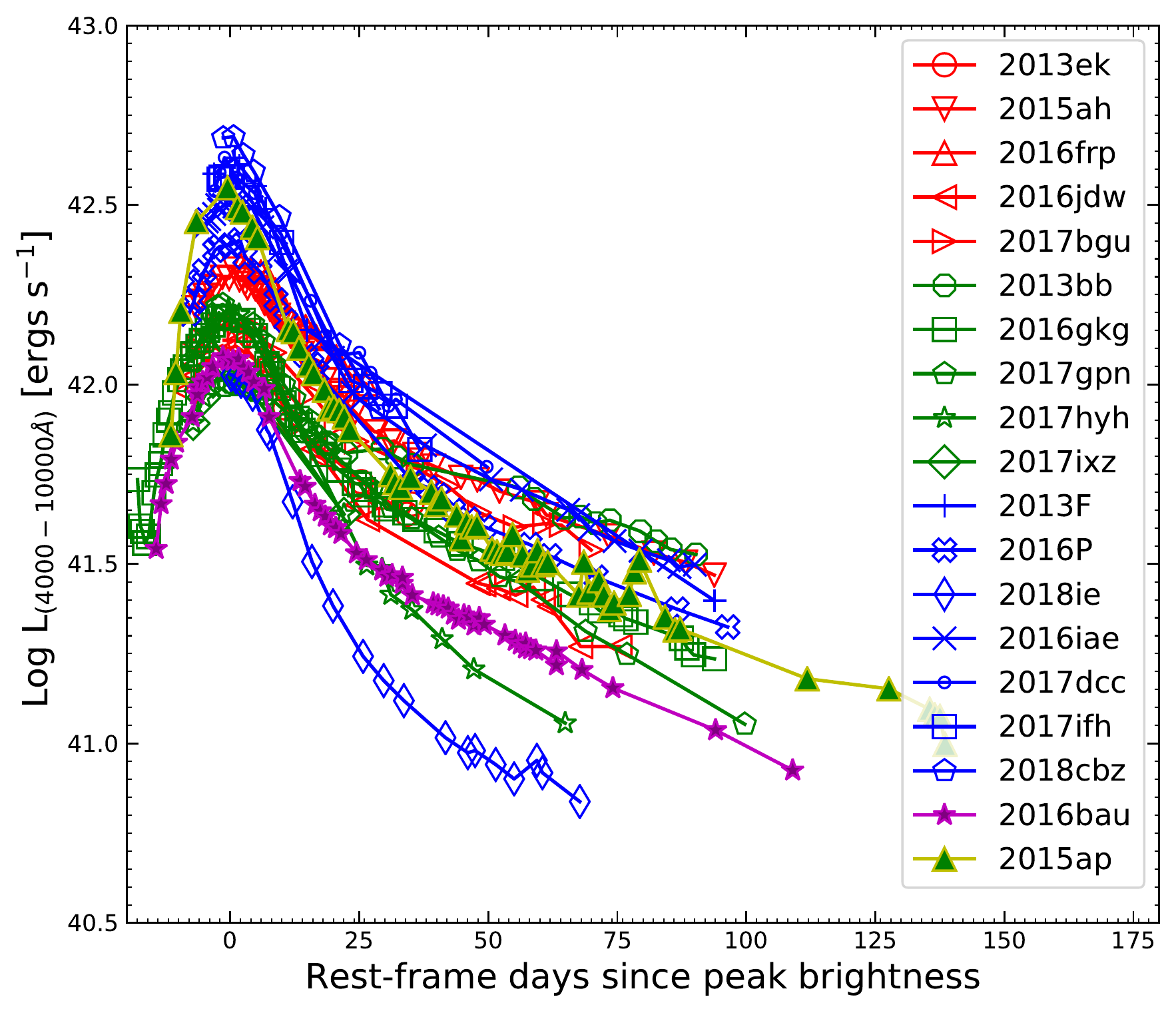}
    \caption{Comparison of quasi-bolometric light curves of SN~2015ap and SN~2016bau, obtained by fitting a blackbody to the SED and integrating the fluxes over the wavelength range of 4000--10,000\,\si{\angstrom}, with that of other stripped-envelope CCSNe. Symbols are colour coded: red, green, and blue correspond to Type Ib, IIb, and Ic SNe, respectively.}
    \label{fig:bol_compare}
\end{figure}

Figure~\ref{fig:bol_compare} shows a comparison of the quasi-bolometric light curve of SN~2015ap with that of other H-stripped CCSNe.
The code {\tt superbol} also provides the  bolometric light curve, including the additional blackbody corrections to the observed quasi-bolometric light curve, by fitting a single blackbody to observed fluxes at a particular epoch and integrating the fluxes trapezoidally for a wavelength range of 100--25,000\,\si{\angstrom}.
The top panel of Figure~\ref{fig:BB_param_SN2015ap} provides the resulting quasi-bolometric and bolometric light curves of SN~2015ap.

\subsubsection{Temperature, radius, and velocity evolution}
\label{tempradvel_SN2015ap}

\begin{figure}
\includegraphics[width=\columnwidth]{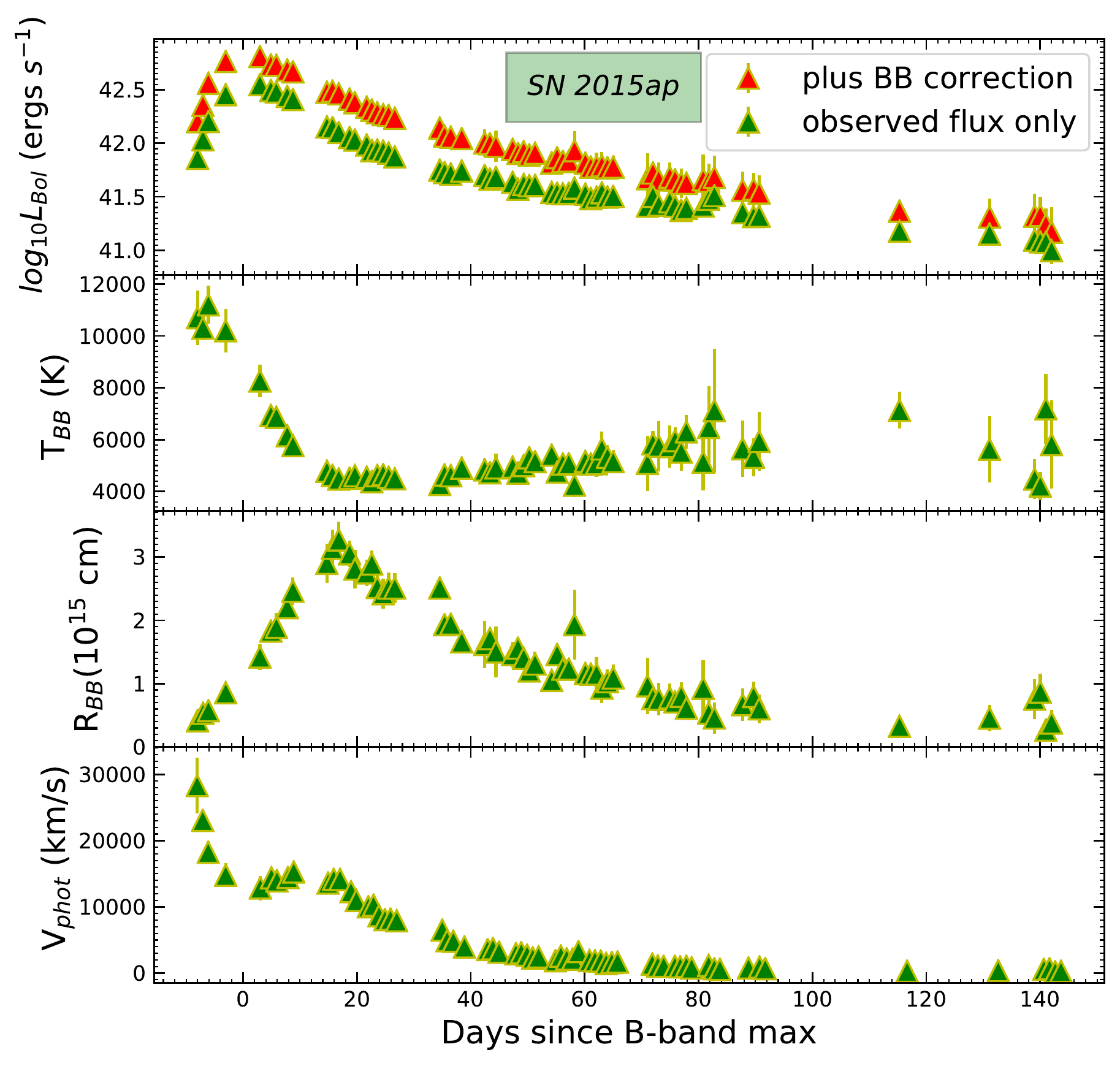}
\caption{The top panel shows the bolometric and quasi-bolometric light curves of SN~2015ap. Second panel: the temperature evolution of SN~2015ap. Third and fourth panels:  the radius and velocity evolutions (respectively) obtained using blackbody fits. }
\label{fig:BB_param_SN2015ap}
\end{figure}

From {\tt superbol}, the photospheric temperature ($T_{\rm BB}$) and radius ($R_{\rm BB}$) evolution of SN~2015ap are also obtained. During the initial phases, the photospheric temperature is high, reaching about 11,400\,K at $-6.18$\,d. Further, as the SN ejecta expand, cooling occurs and the temperature tends to fall, dropping to 4490\,K on +21.66\,d, then remaining nearly constant (Fig.~\ref{fig:BB_param_SN2015ap}, second panel from top). A conventional evolution in radius is also seen. Initially, at an epoch of $-8$\,d, the photospheric radius is $4.14 \times 10^{14}$\,cm. Thereafter, the SN expands and its radius increases, reaching a maximum radius of $3.27 \times 10^{15}$\,cm, beyond which the photosphere seems to recede into the SN ejecta (Fig.~\ref{fig:BB_param_SN2015ap}, third panel from top).
From the  prior knowledge of the explosion epoch and radii at various epochs, we can estimate the photospheric velocity evolution of this SN, using $v_{\rm ph} = R_{\rm BB}/t$, where $t$ is the time since explosion. The bottom panel of Figure~\ref{fig:BB_param_SN2015ap} shows the velocity evolution of SN~2015ap.

From the known value of $t_{\rm exp}$, a rise time ($t_{\rm rise}$) of $14.8\pm2.2$\,d is obtained. The photospheric velocity near maximum light is 9000\,km\,s$^{-1}$ \citep[][]{Prentice2019}. With the known values of $t_{\rm rise}$, the photospheric velocity near maximum light, and a constant opacity ($\kappa$) of 0.07\,cm$^2$\,g$^{-1}$, we also obtain the ejecta mass ($M_{\rm ej}$) and kinetic energy ($E_{\rm ke}$) from the \citet{Arnett1982} model by following Equations (1) and (3) of \citet[][]{Wheeler2015}: $2.2 \pm 0.6\,M_\odot$ and $(1.05\pm0.31) \times 10^{51}$\,erg, respectively. Our derived ejecta mass is slightly higher than that of \citet[][]{Prentice2019}, but less than \citet[][]{Anjasha2020}.
Corresponding to a peak luminosity of $(3.53\pm0.16) \times 10^{42}$\,erg\,s$^{-1}$, the amount of $^{56}$Ni synthesised is $0.14\pm0.02\,M_\odot$, calculated following \citet[][]{Prentice2016}.
 
\subsection{Photometric properties of SN~2016bau}
\label{subsec:Photometric_SN2016bau}

\begin{figure}
	\includegraphics[width=\columnwidth]{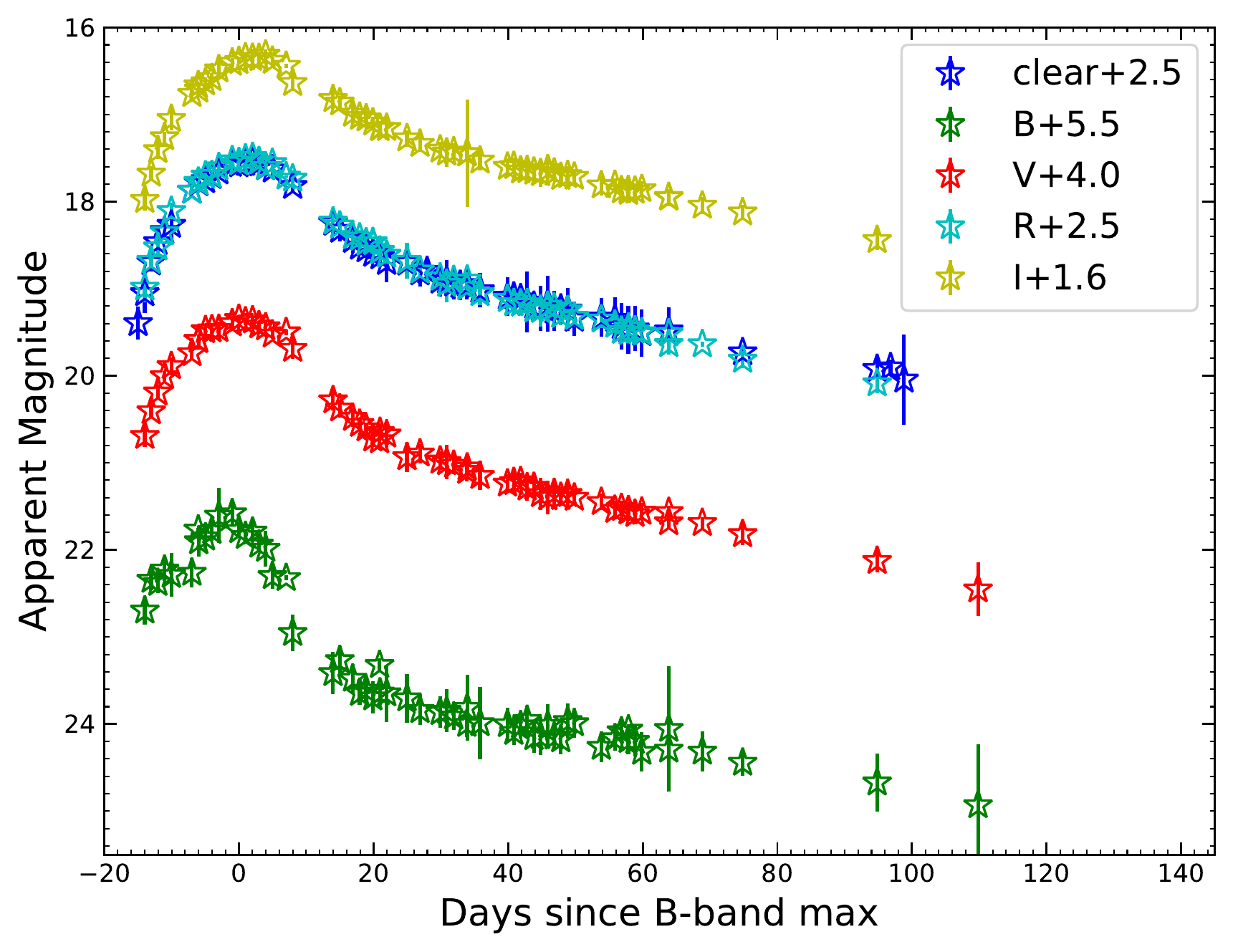}
    \caption{$BVRI$ and Clear filter light curves of SN~2016bau.}
    \label{fig:CBVRI_SN2016bau}
\end{figure}

Following a method similar to that for SN~2015ap, the $B$-band maximum and the explosion epochs of SN~2016bau were determined to be MJD $57477.37 \pm 1.99$ and MJD $57462.54 \pm 0.97$, respectively. Figure~\ref{fig:CBVRI_SN2016bau} shows the $BVRI$ and Clear ($C$) filter light curves of SN~2016bau. Light curves in the shorter-wavelength bands are narrower compared to those in the longer-wavelength bands, following a trend similar to that of SN~2015ap. Also, the $C$ filter almost exactly replicates the $R$-band light curve, as expected \citep{Li2003}.

\subsubsection{Colour evolution and extinction correction}

\begin{figure}	\includegraphics[width=\columnwidth]{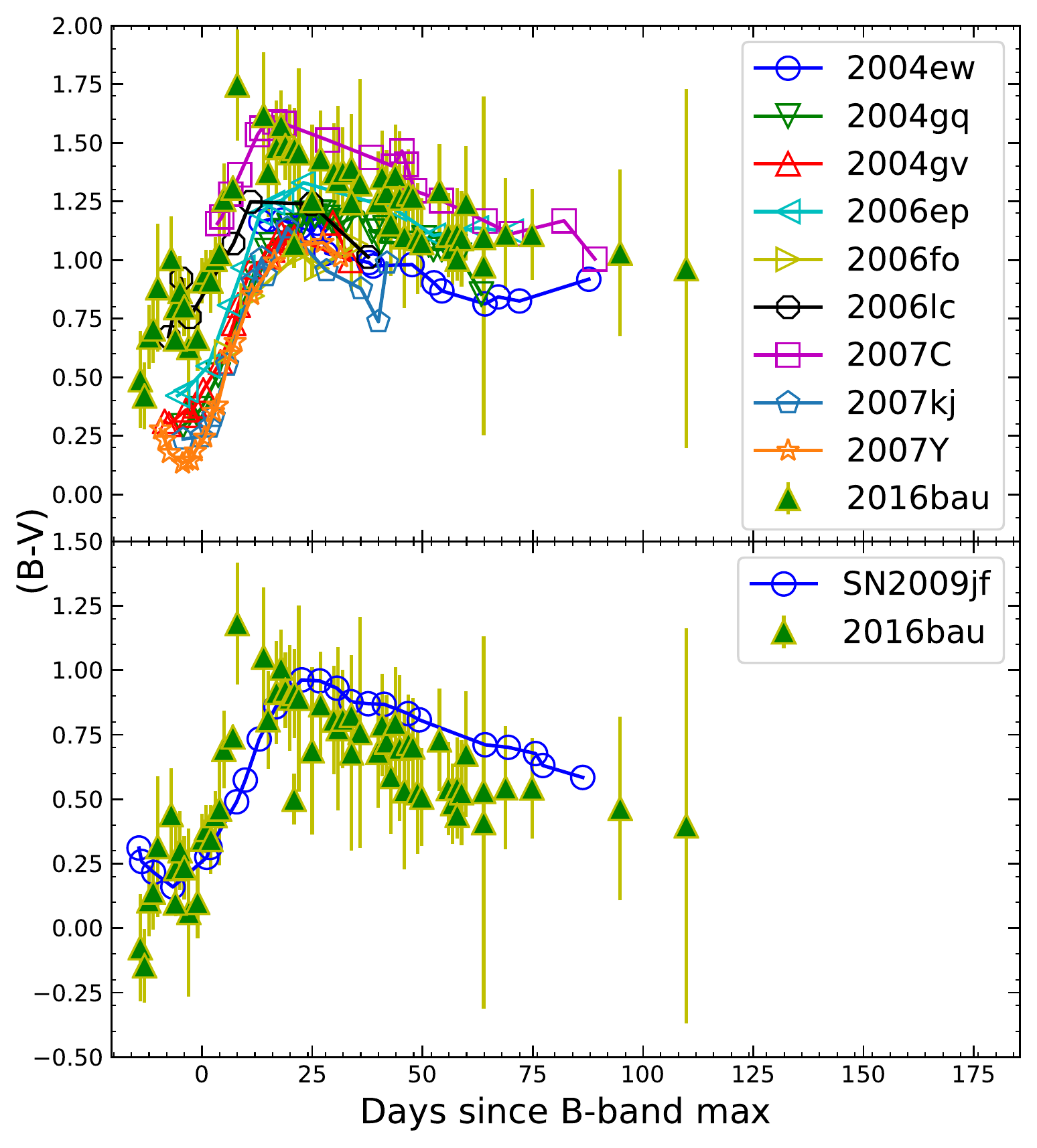}
    \caption{The top panel shows a comparison of the $(B-V)$ colour of SN~2016bau with that of other SNe~Ib (all corrected for MW extinction). The data for the other SNe are taken from \citet[][]{Stritzinger2018}. The bottom panel shows the $(B-V)$ colour curves of SN~2016bau and SN~2009jf, both corrected for MW extinction. To match the SN~2009jf colour curve, a shift of 0.566\,mag is required for the SN~2016bau colour curve.}
    \label{fig:color_curve_2016bau}
\end{figure}

Similar to SN~2015ap, we corrected for MW extinction using NED following \citet[][]{Schlafly2011}. In the direction of SN~2016bau, the Galactic extinction for the $B$, $V$, $R$, and $I$ bands is 0.060, 0.045, 0.036, and 0.025\,mag, respectively. The top panel of Figure~\ref{fig:color_curve_2016bau} shows a comparison of the $(B-V)$ colour of SN~2016bau with that of other SNe~Ib. SN~2016bau seems to be heavily reddened and lies above nearly all of the other SNe~Ib. To correct for the host-galaxy extinction, we made use of the colour curve of SN~2009jf. We calculated the differences in the colours of the two SNe and took their weighted mean. In this calculation, we made use of data only in the range 0 to +20\,d for the reasons mentioned by \citet[][]{Stritzinger2018}. The resulting host-galaxy extinction is $E(B-V)_{\rm host} = 0.566\pm0.046$\,mag. Thus, in order to match the SN~2009jf $(B-V)$ colour curve, we need to shift the MW-corrected $(B-V)$ colour curve of SN~2016bau downward by 0.566\,mag, as shown in  the bottom panel of Figure~\ref{fig:color_curve_2016bau}.

\subsubsection{Quasi-bolometric and bolometric light curves}

To obtain the quasi-bolometric and bolometric light curves of SN~2016bau, we used {\tt superbol}, as in the case of SN~2015ap. The extinction-corrected  $B$, $V$, $R$, and $I$ data were given as input to {\tt superbol}. The top panel of Figure~\ref{fig:BB_param_SN2016bau} shows the quasi-bolometric and bolometric light curves of SN~2016bau. Figure~\ref{fig:bol_compare} shows the comparison of the quasi-bolometric light curve of SN~2016bau obtained by integrating the flux over the wavelength range 4000--10,000\,\AA\ with other H-stripped CCSNe from \citet[][]{Prentice2019}. Here, SN~2016bau also seems to lie on the moderately bright end.   
  
\subsubsection{Temperature, radius, and velocity evolution}
\label{tempradvel_SN2016bau}

\begin{figure}
\includegraphics[width=\columnwidth]{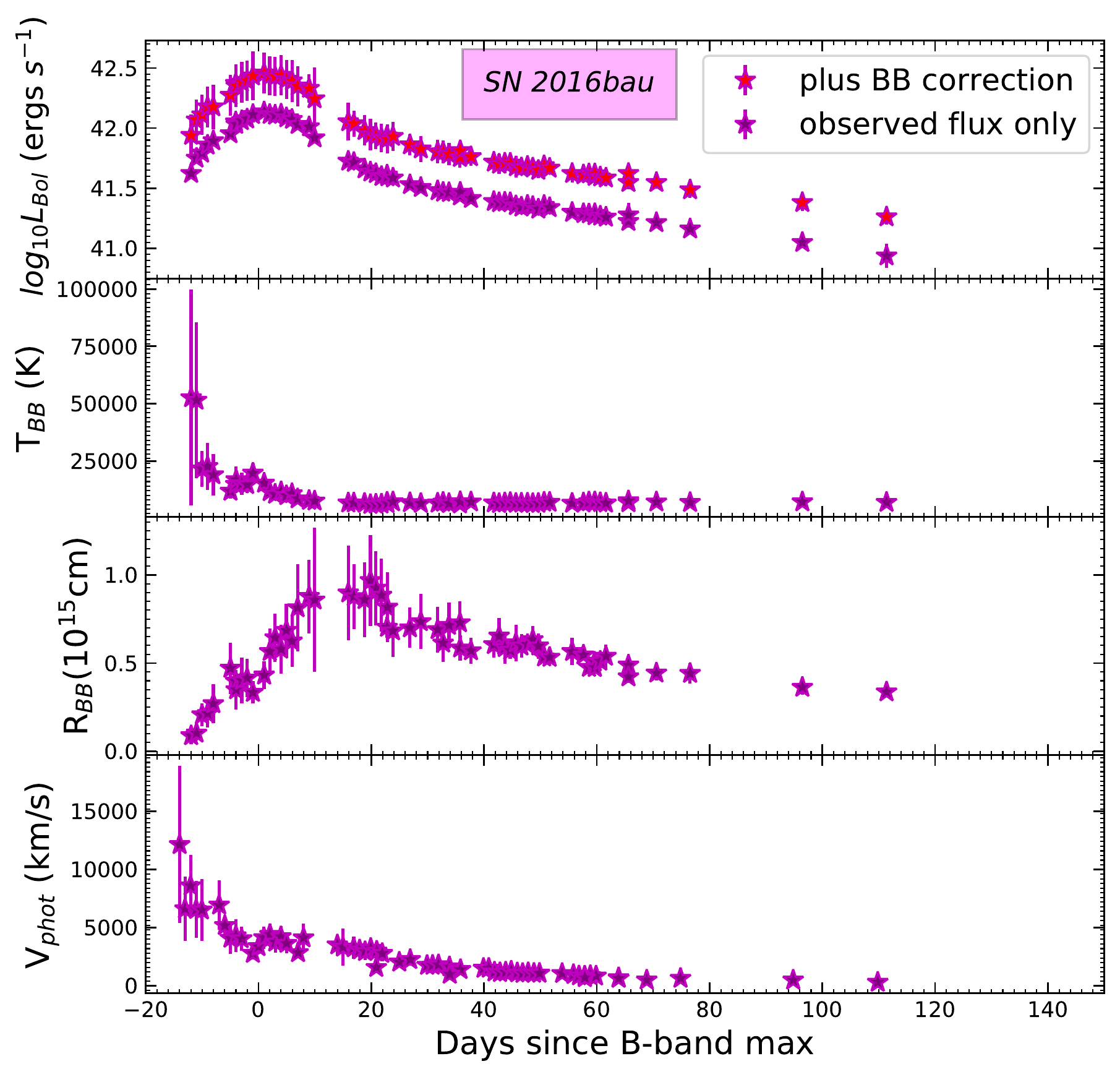}
\caption{The top panel shows the bolometric and quasi-bolometric light curves of SN~2016bau. Second panel: the temperature evolution of SN~2016bau. Third and fourth panels:  the radius and velocity evolution (respectively) obtained using blackbody fits.}
\label{fig:BB_param_SN2016bau}
\end{figure}

Figure~\ref{fig:BB_param_SN2016bau} also shows the evolution of the photospheric temperature ($T_{\rm BB}$) and radius ($R_{\rm BB}$) of SN~2016bau, obtained using {\tt superbol}. During the initial phases, the photospheric temperature is very high, reaching about 17,000\,K near 0\,d. Thereafter, as the SN ejecta expand, cooling occurs and the temperature falls, reaching 6000\,K at around +20\,d, then remaining nearly constant (Fig.~\ref{fig:BB_param_SN2016bau}, second panel from top). A conventional evolution in radius is also seen (Fig.~\ref{fig:BB_param_SN2016bau}, third panel from top). Initially, at an epoch of around $-6$\,d, the photospheric radius is $0.28 \times 10^{15}$\,cm. Following this, the supernova expands and its radius increases, reaching a maximum radius of $0.96 \times 10^{15}$\,cm, beyond which the photosphere seems to recede within the SN ejecta.

From the prior knowledge of explosion epoch and radii at various epochs, we estimate the photospheric velocity evolution of this SN in the same way as for SN~2015ap. The bottom panel of Figure~\ref{fig:BB_param_SN2016bau} shows the velocity evolution of SN~2016bau.
   
From the prior knowledge of $t_{\rm exp}$, a rise time ($t_{\rm rise}$) of $17.09 \pm 1.29$\,d is obtained. The photospheric velocity near maximum light is $\sim 5000$\,km\,s$^{-1}$, obtained from a blackbody fit. With the known values of $t_{\rm rise}$, the photospheric velocity near maximum light, and a constant opacity ($\kappa$) 0.07\,cm$^2$\,g$^{-1}$, we obtain the ejecta mass ($M_{\rm ej}$) and kinetic energy ($E_{\rm ke}$) following Equation (1) and Equation (3) of \citet[][]{Wheeler2015}; the results are $1.6 \pm 0.3\,M_\odot$ and $(0.24 \pm 0.04) \times 10^{51}$\,erg, respectively. Following \citet[][]{Prentice2016}, an amount of $0.055 \pm 0.006\,M_\odot$ of $^{56}$Ni is synthesised, corresponding to a peak luminosity of $(1.19 \pm 0.08) \times 10^{42}$\,erg\,s$^{-1}$.

\section{Spectral studies of SN~2015ap and SN~2016bau}
\label{sec:Spectral}
In this section, we discuss spectral features of SN~2015ap and SN~2016bau and further compare their properties with other similar SNe. We modelled the spectra of these two SNe at different epochs using {\tt SYN++} \citep[][]{Branch2007, Thomas2011} and performed the spectral matching of the 12, 13, and 17\,$M_{\odot}$ spectral models given by \citet[][]{Jerkstrand2015} with the spectra of SN~2015ap and SN~2016bau at phases around 100\,d past explosion. In this section, we also estimate the velocities of various lines present in the spectra, using their absorption troughs.

\subsection{Spectral properties of SN~2015ap}
\label{subsec:Spec_evol_SN2015ap}

\begin{figure}
\includegraphics[width=\columnwidth]{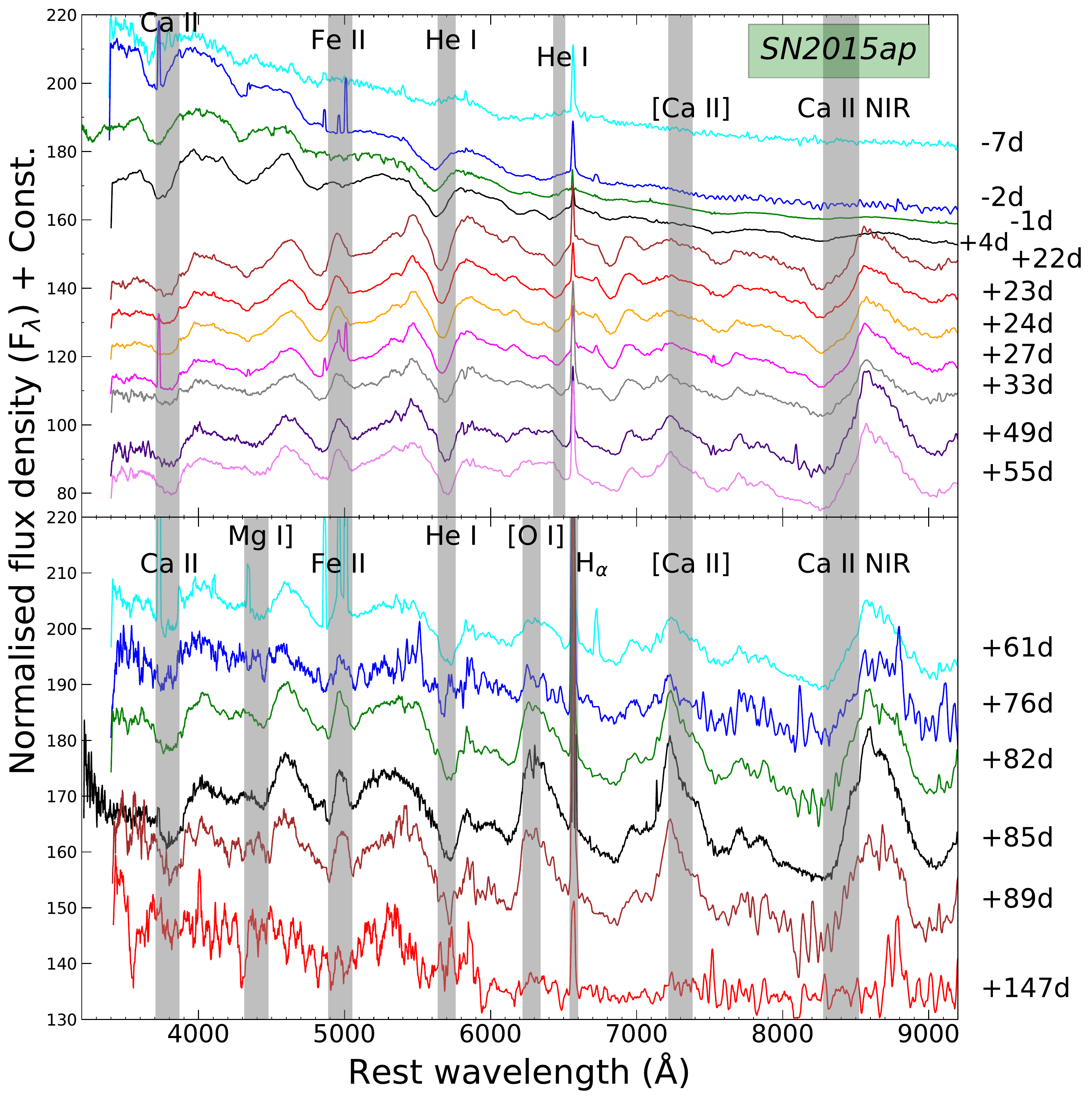}
\caption{Top panel: the early-phase spectra up to +55\,d. Bottom panel: the spectral evolution beyond +55\,d since $B_{\rm max}$ of SN~2015ap.}
\label{fig:spectral_evol_SN2015ap}
\end{figure}

The top and bottom panels of Figure~\ref{fig:spectral_evol_SN2015ap} show the early phases ($-7$\,d to +55\,d) and late phases (+61\,d to +147\,d) spectral evolution, respectively. It is quite evident from the top panel of  Figure~\ref{fig:spectral_evol_SN2015ap} that initially SN~2015ap shows broad-lined features (at $-7$\,d and $-2$\,d), which later evolve to spectra of a normal Type Ib SN. The characteristic He~I line at 5876\,\si{\angstrom}  of a typical SN~Ib is clearly seen in the spectra of SN~2015ap. We also see unambiguous He~I features at 6678\,\si{\angstrom} and 7065\,\si{\angstrom}, but these two lines are not as prominent as the one at 5876\,\si{\angstrom}. In the very early phases (up to +4\,d), the He~I line at 7065\,\si{\angstrom} is hard to identify. We see that the He~I absorption at 5876\,\si{\angstrom} is stronger than any other absorption features, and it is present in every spectrum up to +147\,d. The Fe~II feature near 5169\,\si{\angstrom} is very hard to be identified, and it seems to be highly blended with He~I 5016\,\si{\angstrom}. 

The spectral evolution also shows the Ca~II near-infrared (NIR) feature, which is almost absent in the very early phases ($-7$\,d to +4\,d), but starts to develop very strongly from +22\,d onward. It is so strong that each spectrum on and after +22\,d shows it, even the lowest-SNR spectrum at +147\,d. The forbidden [Ca~II] feature near 7300\,\si{\angstrom} is almost absent at very early phases ($-7$\,d to +4\,d), then begins to develop very obviously in the spectra from +22\,d, and is present up to +89\,d. This [Ca~II] feature can also get blended with [O~II] emission at 7320\,\si{\angstrom} and 7330\,\si{\angstrom}. It almost disappears in the spectrum at +147\,d, though its absence may not be real owing to the very poor SNR of that spectrum. We also see the Ca~II~H\&K feature near 3934\,\si{\angstrom}, which is present in every spectrum.
From the bottom panel of Figure~\ref{fig:spectral_evol_SN2015ap}, we see that as the spectra evolve, the absorption features of different lines tend to disappear and the emission features become more prominent. We see a very weak semiforbidden Mg~I] line, at 4571\,\si{\angstrom}. However we can clearly see the forbidden emission lines of [O~I] and [Ca~II], which indicates the onset of the nebular phase.

\subsubsection{Spectral comparison}
\label{subsec:spec_com_SN2015ap}

To investigate the spectroscopic behaviour of SN~2015ap, we compare its spectral features with those of other well-studied SNe~Ib such as SN~2004gq \citep[][]{Modjaz2014}, SN~2008D \citep[][]{Modjaz2014}, SN~2012au \citep[][]{Pandey2020}, SN~2009jf \citep[][]{Sahu2011, Modjaz2014}, iPTF13bvn \citep[][]{Srivastav2014a}, SN~2007uy \citep[][]{Modjaz2014, Milisavljevic2010}, SN~2007gr \citep[][]{Valenti2008, Modjaz2014}, and SN~2005bf \citep[][]{Modjaz2014}.

\begin{figure}
\includegraphics[width=\columnwidth]{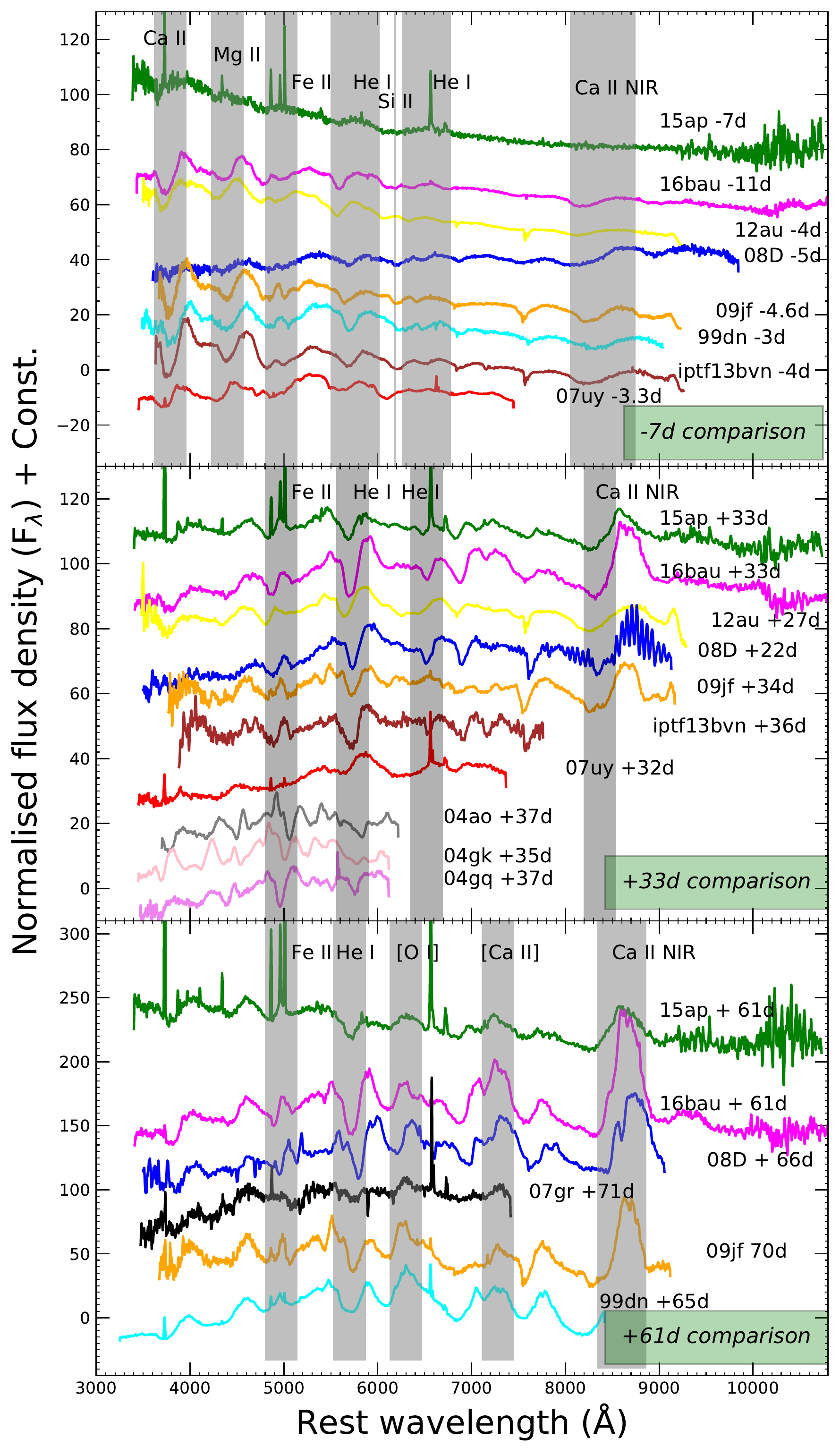}
\caption{Spectral comparison of the spectra of SN~2015ap and SN~2016bau at epochs $-7$\,d (top panel), +33\,d (middle panel), and +61\,d (bottom panel). Various important SN~Ib features have been compared with those of other similar-type SNe.}
\label{fig:spectral_comparison}
\end{figure}

The top panel of Figure~\ref{fig:spectral_comparison} shows the early phase ($-7$\,d) spectral comparison of SN~2015ap with these well-studied SNe~Ib. The spectral features of SN~2015ap look similar to those of SN~2008D and SN~2012au, compared to other SNe~Ib. We can see that the Ca~II~H\&K feature, the Mg~II feature, and the He~I P~Cygni profile match very well with those of SN~2008D, while in other SNe in the comparison sample, these features are much more developed. The blue end of the spectrum matches nicely with SN~2008D, while the redder part is featureless and much closer to SN~2012au and SN~2005bf. The He~I 5876\,\si{\angstrom} feature of SN~2015ap is completely different from that of SN~2007uy, but resembles that of SN~2008D and seems to be less evolved compared to that of SN~2004gq, SN~2012au, SN~2009jf, and iptf13bvn. The Fe~II profile in SN~2015ap is hard to detect, which may be due to a very high initial optical opacity. We try to estimate the velocities using the various absorption features. As the spectrum at this epoch is continuum dominated, only a few absorption features were visible. The velocity estimated using He~I absorption lines  is $\sim$\,14,100\,km\,s$^{-1}$.

Further, we compare the +33\,d spectrum of SN~2015ap with our comparison sample (middle panel of Figure~\ref{fig:spectral_comparison}). At this epoch, the spectrum of SN~2015ap shows various P~Cygni profiles. We see that it almost exactly replicates the +22\,d spectrum of SN~2008D. The blended He~I and Fe~II profiles near 5016\,\si{\angstrom} well match those of other SNe~Ib, except for SN~2007uy, where the peak is almost absent, and for SN~2005bf, where it is double peaked. The He~I feature at 5876\,\si{\angstrom} matches well with other SNe~Ib except for the cases of SN~2007uy and SN~2005bf, where the absorption features seem to be much broader. Here, once again, SN~2007uy seems to match much less and SN~2008D seems to best match the +22\,d spectrum of SN~2015ap. The velocities estimated using the Ca~II NIR triplet and He~I absorption features are $\sim$\,7800\,km\,s$^{-1}$ and 10000\,km\,s$^{-1}$, respectively.

For a much clear comparison, we compared the +61\,d  spectrum of SN~2015ap (bottom panel of Fig.~\ref{fig:spectral_comparison}) with spectra of other well-studied SNe~Ib. At this epoch also, the spectrum is still much closer to that of SN~2008D compared to other SNe~Ib. The blended Fe~II and He~I  feature near 5016\,\si{\angstrom} shows an almost similar profile to that of SN~2008D. However, the 5876\,\si{\angstrom} He~I profile is narrower in SN~2015ap compared to SN~2008D. We can see the onset of the appearance of the [O~I] line in SN~2015ap, which differs from SN~2008D. The He~I profile is well matched in the cases of SN~1999dn and SN~2009jf, but the Ca~II NIR triplet of these SNe differs from that of SN~2015ap. Here the velocities estimated using the Ca~II NIR triplet and He~I absorption features are $\sim$\,6900\,km\,s$^{-1}$ and 8100\, km\,s$^{-1}$, respectively.

\subsubsection{Spectral modelling}
\label{subsec:syn_SN2015ap}
\begin{figure}
	\includegraphics[width=\columnwidth]{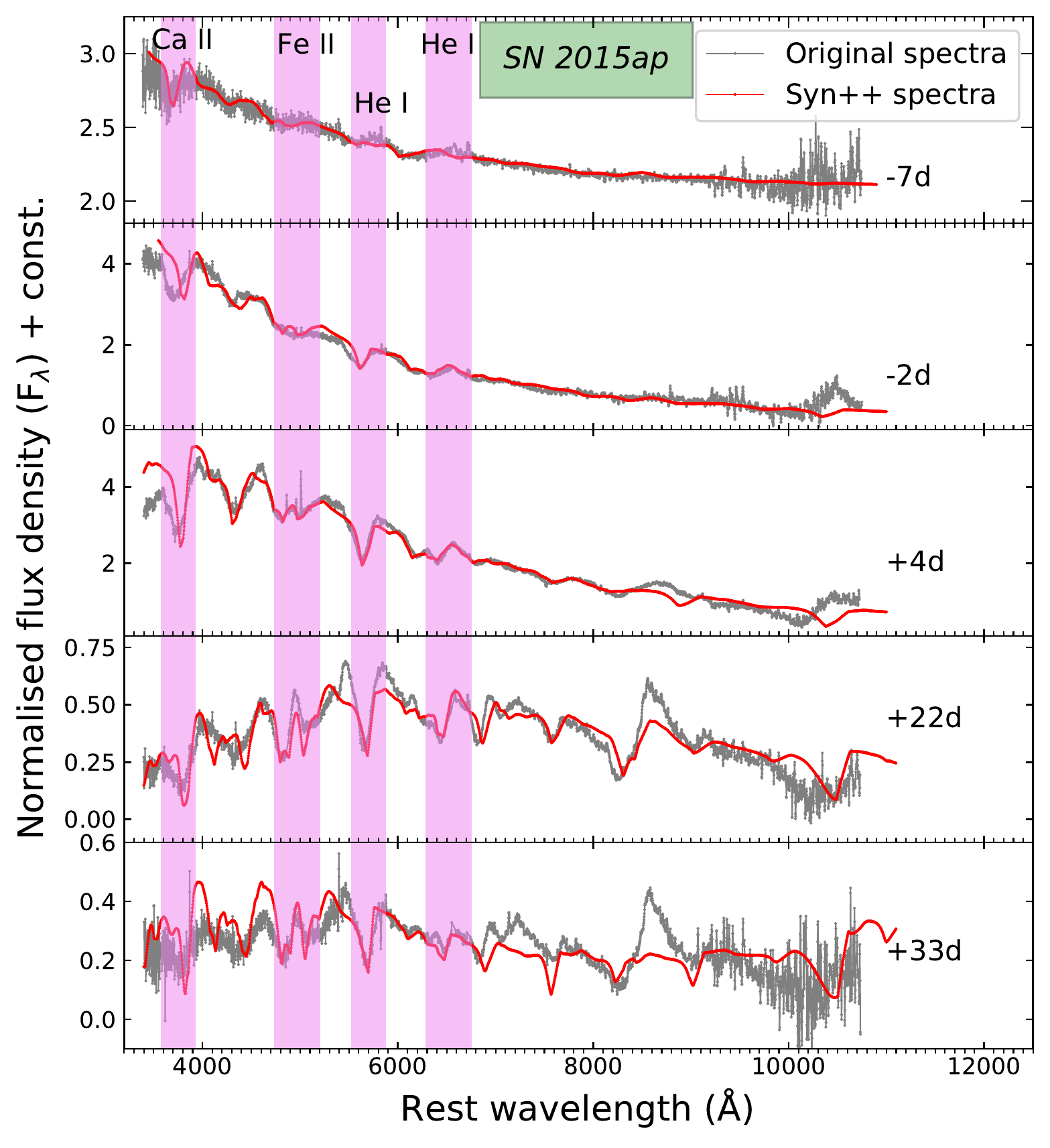}
    \caption{{\tt SYN++} modelling of the spectra of SN~2015ap at epochs $-7$\,d, -2\,d, +4\,d, +22\,d, and +33\,d. Prominent He~I features could be produced nicely.}
    \label{fig:syn_SN2015ap}
\end{figure}

After confidently identifying the features present in the spectra of SN~2015ap and comparing them with those of other well-studied SNe, we tried to model a few spectra at various epochs using {\tt SYN++}. The top panel in Figure~\ref{fig:syn_SN2015ap} shows the early-phase ($-7$\,d) spectrum of SN~2015ap. This spectrum displays weak and broad P~Cygni profiles of He~I, Ca~II~H\&K, and the Ca~II~NIR triplet, and also some blended features of Fe~II. We also show the best-matching synthetic spectrum, generated by {\tt SYN++}. The absorption features due to Ca~II~H\&K, Ca~II~NIR triplet, He~I, and Fe~II multiplets are easily reproduced. The photospheric velocity and blackbody temperature associated with the best-fit spectrum are 12,000\,km\,s$^{-1}$  and 12,000\,K, respectively. We perform {\tt SYN++} matching to four additional spectra from epochs $-2$\,d to +33\,d (Fig.~\ref{fig:syn_SN2015ap}). With the passage of time, the SN expands, cools gradually, and its expansion velocity decreases slowly, so we see a gradual decrease in the values of these fitted parameters. The photospheric velocities during the phase of $-7$\,d to +33\,d vary from 13,000\,km\,s$^{-1}$ to 6800\,km\,s$^{-1}$, and the blackbody temperature varies from 12,000\,K to 4500\,K, which are in good agreement with those obtained photometrically from blackbody fits. Owing to the local thermodynamic equilibrium (LTE) approximation, {\tt SYN++} does not work well for the later epochs and cannot be used to fit the spectra.

To get an estimate of the progenitor mass, we used the +98.75\,d post-explosion spectrum of SN~2015ap and plotted it along with the 12, 13, and 17\,$M_{\odot}$ model spectra from \citet[][]{Jerkstrand2015} at 100\,d, after scaling by a factor of exp$(-2 \times \Delta t /111.4)$ \citep[][]{Jerkstrand2015}, where $\Delta t$ = 1.25, is the time difference between the epoch of the model spectrum and the epoch of the observed spectrum. We can see that the 12\,$M_{\odot}$ and 17\,$M_{\odot}$ model spectra seem to reproduce the observed spectrum of SN~2015ap. The 12\,$M_{\odot}$ spectrum very nicely matches the observed spectrum throughout the entire wavelength range, while the 17\,$M_{\odot}$ spectrum slightly overproduces the flux near the Ca~II~NIR triplet close to 8500\,\si{\angstrom}. However, the 13\,$M_{\odot}$ model spectrum fails to explain the observed fluxes throughout the entire wavelength range. Thus, based on this analysis, a range of 12--17\,$M_{\odot}$ is expected for the possible progenitor mass of SN~2015ap, in agreement with that described by \citet[][]{Anjasha2020}.

\begin{figure}
\includegraphics[width=\columnwidth]{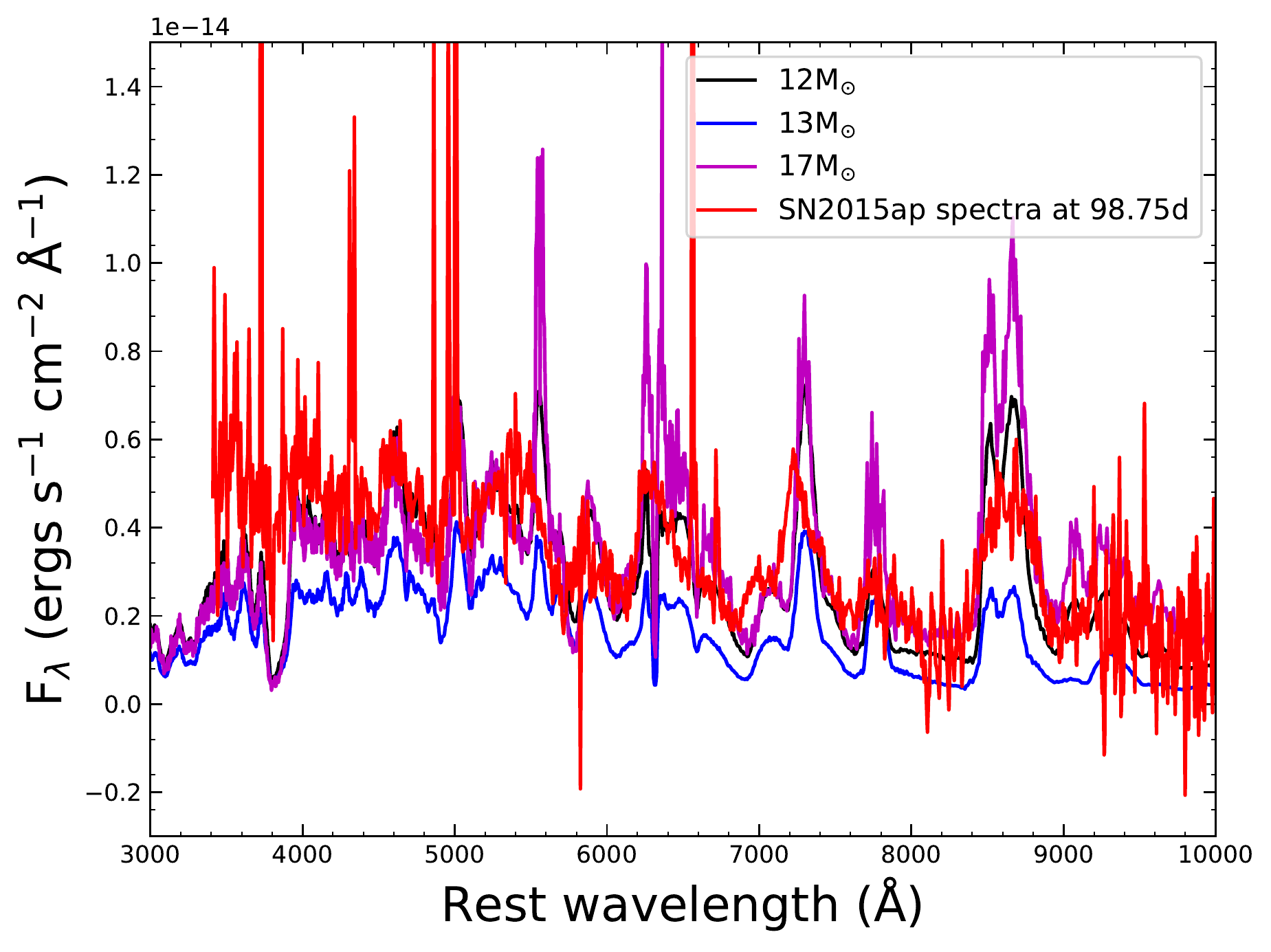}
\caption{The $t = +98.75$\,d  spectrum of SN~2015ap plotted along with the 12, 13, and 17\,$M_{\odot}$ models from \citet[][]{Jerkstrand2015} at 100\,d scaled with an exponential factor exp$(-2 \times 1.25/111.4)$. The 12\,$M_{\odot}$ and 17\,$M_{\odot}$ models seem to best match the observed spectrum of SN~2015ap.}
\label{fig:jerkstrand_SN2015ap}
\end{figure} 

\begin{figure*}
\centering
\includegraphics[width=\textwidth]{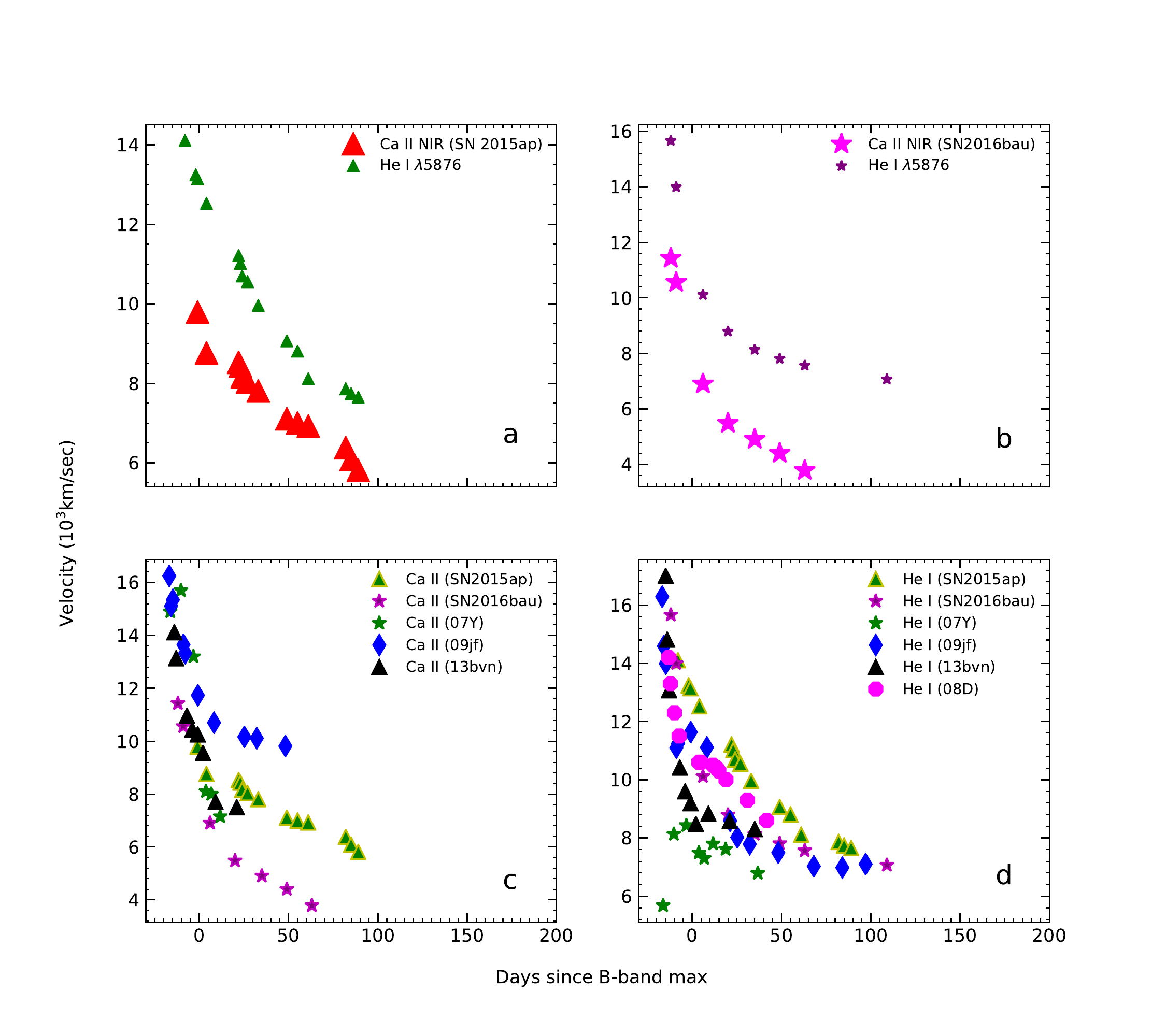}
\caption{(a) The temporal evolution of velocities of He~I 5876\,\si{\angstrom} and the Ca~II NIR triplet for SN~2015ap. (b) Same as (a) but for SN 2016bau. The bottom two panels show the comparison of these line velocities of SN~2015ap and SN~2016bau with SNe 2007Y \citep[][]{Stritzinger2009}, 2009jf \citep[][]{Sahu2011}, iPTF13bvn \citep[][]{Srivastav2014a}, and 2008D \citep[][]{Modjaz2014}. }
\label{fig:vel_evolve}
\end{figure*}

\subsubsection{Velocity evolution of various lines of SN~2015ap}
\label{subsec:Vel_evol_SN2015ap}

We used the blue--shifted absorption minima of P~Cygni profiles and the special-relativistic Doppler formula \citep[][]{Sher1968} to obtain the velocities of several lines. Figure~\ref{fig:vel_evolve}$a$ shows the He~I and the Ca~II NIR triplet velocity evolution of SN~2015ap. In the initial few days, the line velocities tend to decrease rapidly.
At an epoch of +24\,d, the velocities estimated using He~I 5876\,\si{\angstrom} and the Ca~II NIR triplet are $\sim$\,10,600\,km\,s$^{-1}$ and 8200\,km\,s$^{-1}$, respectively. The velocities estimated using these two lines drop to $\sim$\,9100\,km\,s$^{-1}$ and 7100\,km\,s$^{-1}$ (respectively) at an epoch of +49\,d. In the late phases, the velocities decline gradually.

Figure~\ref{fig:vel_evolve}$c,d$, show comparisons of velocities obtained using the Ca~II NIR triplet and He~I 5876\,\si{\angstrom} features with other well-studied SNe. Initially the Ca~II NIR line velocity of SN~2015ap evolves in a manner similar to iptf13bvn, thereafter it decays slowly, attaining a velocity of $\sim$\,7600\,km\,s$^{-1}$ at +89\,d. We see that the ejecta velocity of SN~2015ap obtained using He~I 5876\,\si{\angstrom} is higher than that of other SNe~Ib but closer to SN~2008D.   

\subsection{Spectral properties of SN~2016bau}
\label{subsec:Spec_evol_SN2016bau}

\begin{figure}
\includegraphics[width=\columnwidth]{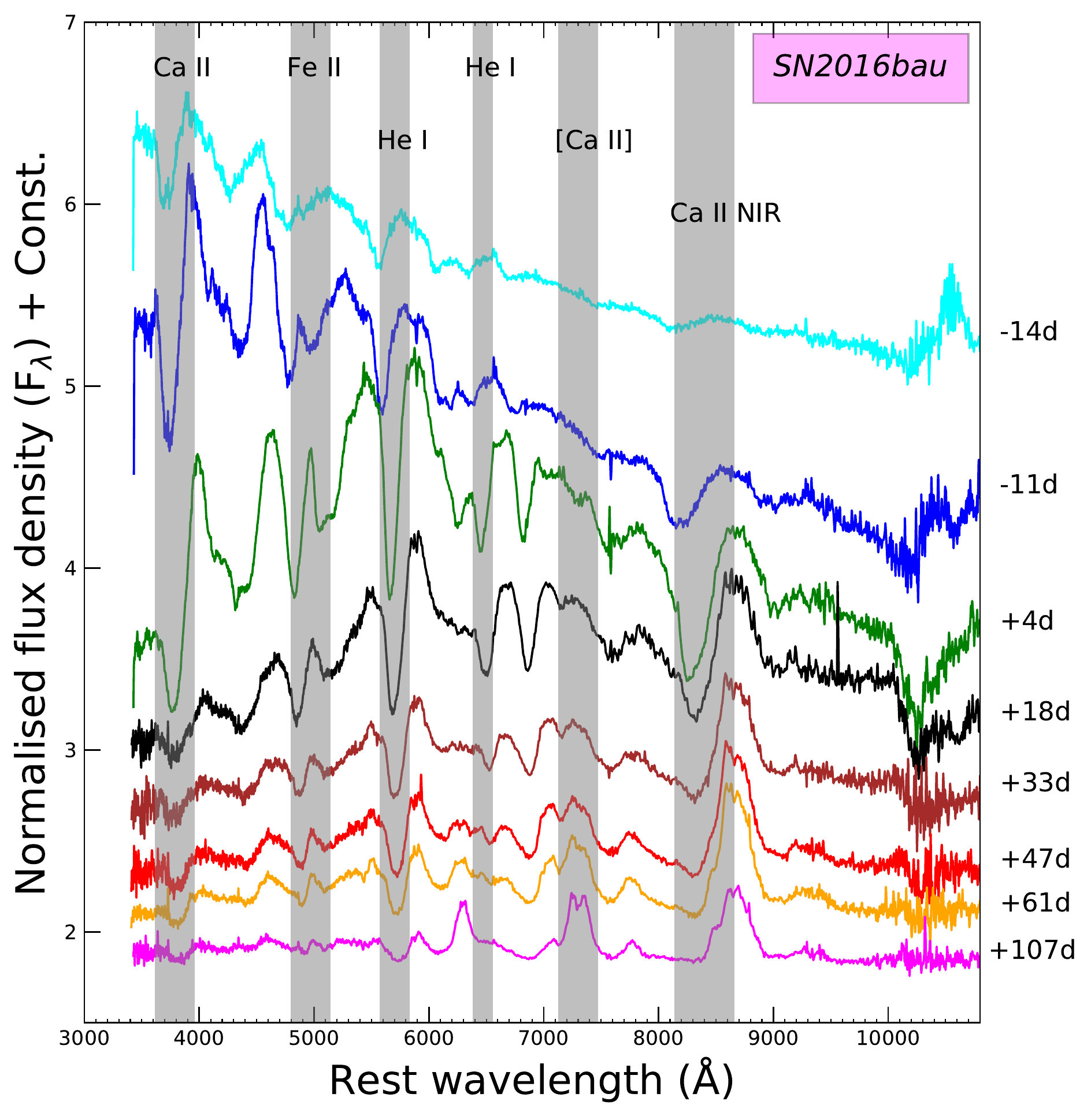}
\caption{Various line identifications in the overall spectral evolution of SN~2016bau. Strong He~I features along with other important lines are indicated.}
\label{fig:spec_evol_SN2016bau}
\end{figure}

Figure~\ref{fig:spec_evol_SN2016bau} shows the spectral evolution of SN~2016bau for a period of $-14$\,d to +107\,d. In the first spectrum ($-14$\,d), we see well-developed He~I features (5876, 6678, and 7065\,\si{\angstrom}). Also, the spectrum on this particular epoch displays a weak and broad Ca~II~NIR feature, but with the passage of time (beyond $\sim +33$\,d), prominent Ca~II NIR features start to appear. We also see that in the initial phases there is a strong Ca~II~H $\&$ K feature, but it becomes progressively weaker at later phases. As the phase approaches the date of $B$-band maximum brightness, the He~I features start to appear much more strongly, as seen in the spectra at $-11$\,d, +4\,d, and +18\,d. Beyond $-11$\,d, the longer wavelengths exhibit strong Ca~II~NIR features. The spectra also display very strong P~Cygni profiles of He~I at 5016\,\si{\angstrom}  after $t =  -11$\,d. The emergence of very strong features of He~I in the early-time spectra confirms this SN to be of Type Ib.

\subsubsection{Spectral comparison}
\label{subsec:spec_com_SN2016bau}

To investigate the spectroscopic behaviour of SN~2016bau, we have compared its spectral features with those of other well-studied SNe~Ib. We used a sample similar to that for SN~2015ap.
The top panel of Figure~\ref{fig:spectral_comparison} shows the early-phase ($-11$\,d) spectral comparison of SN~2016bau with other well-studied H-stripped CCSNe. The spectral features of SN~2016bau look much more similar  to those of SN~2012au, SN~2009jf, and SN~1999dn compared to other SNe. we can see that the Ca~II~ H\&K feature, the Mg~II feature, and the He~I P~Cygni profile of SN~2016bau match very well those of SN~2012au, SN~2009jf, and SN~1999dn. The He~I 5876\,\si{\angstrom} feature of SN~2016bau is completely different from that of SN~2007uy, SN~2015ap, and SN2008D. As with SN~2015ap, the blended Fe~II profile in SN~2016bau is hard to detect, which may be due to a very high initial optical opacity. The spectrum at this epoch has nicely developed He~I features. The velocity estimated using the He~I 5876\,\si{\angstrom} absorption line is $\sim$\,15,600\,km\,s$^{-1}$.

The middle panel of Figure~\ref{fig:spectral_comparison} shows the +33\,d spectral comparison of SN~2016bau with other well-studied SNe~Ib. The spectrum at this epoch contains many P~Cygni profiles of various lines. These spectral features look most similar to those of SN~2008D and SN~2015ap, compared to other SNe~Ib. We see that the Ca~II NIR feature, the Fe~II feature, and the He~I P~Cygni profile match very well those of SN~2008D and SN~2015ap, compared to other SNe. SN~2009jf also seems to match nicely in the redder part of the spectrum. The He~I 5876\,\si{\angstrom} feature of SN~2016bau is completely different from those of SN~2007uy, SN~2005bf, SN~2004gq, and SN~2007gr, but resembles those of SN~2008D, SN~2015ap, and SN2009jf. The He~I 5876\,\si{\angstrom}  profile of SN~2016bau seems to be more asymmetric and narrower than that of iptf13bvn. At this epoch, the velocities estimated using the absorption features of He~I and the Ca~II NIR triplet are $\sim$\,8100\,km\,s$^{-1}$ and 4900\,km\,s$^{-1}$, respectively.

For much clearer comparisons, we also compared the +61\,d spectrum of SN~2016bau (bottom panel of Figure.~\ref{fig:spectral_comparison}) with other well-studied H-stripped CCSNe spectra. At this epoch, the bluer part of the spectrum is much closer to SN~2008D compared to other SNe~Ib. Here the blended Fe~II and He~I features near 5016\,\si{\angstrom} show profiles nearly similar to those of SN~2008D and SN~2009jf. However, the 5876\,\si{\angstrom} He~I profile is broader in SN~2016bau compared with SN~2008D. We can see that the onset of the appearance of [O~I] in SN~2016bau is slightly different from that of SN~2008D, while it matches nicely that of SN~2009jf. The He~I profile is well matched in the case of SN~2008D. Here the velocities estimated using the Ca~II NIR and He~I absorption features are $\sim$\,3700\,km\,s$^{-1}$ and 7600\,km\,s$^{-1}$, respectively.

\subsubsection{Spectral modelling}
\label{subsec:syn_SN2016bau}

\begin{figure}
	\includegraphics[width=\columnwidth]{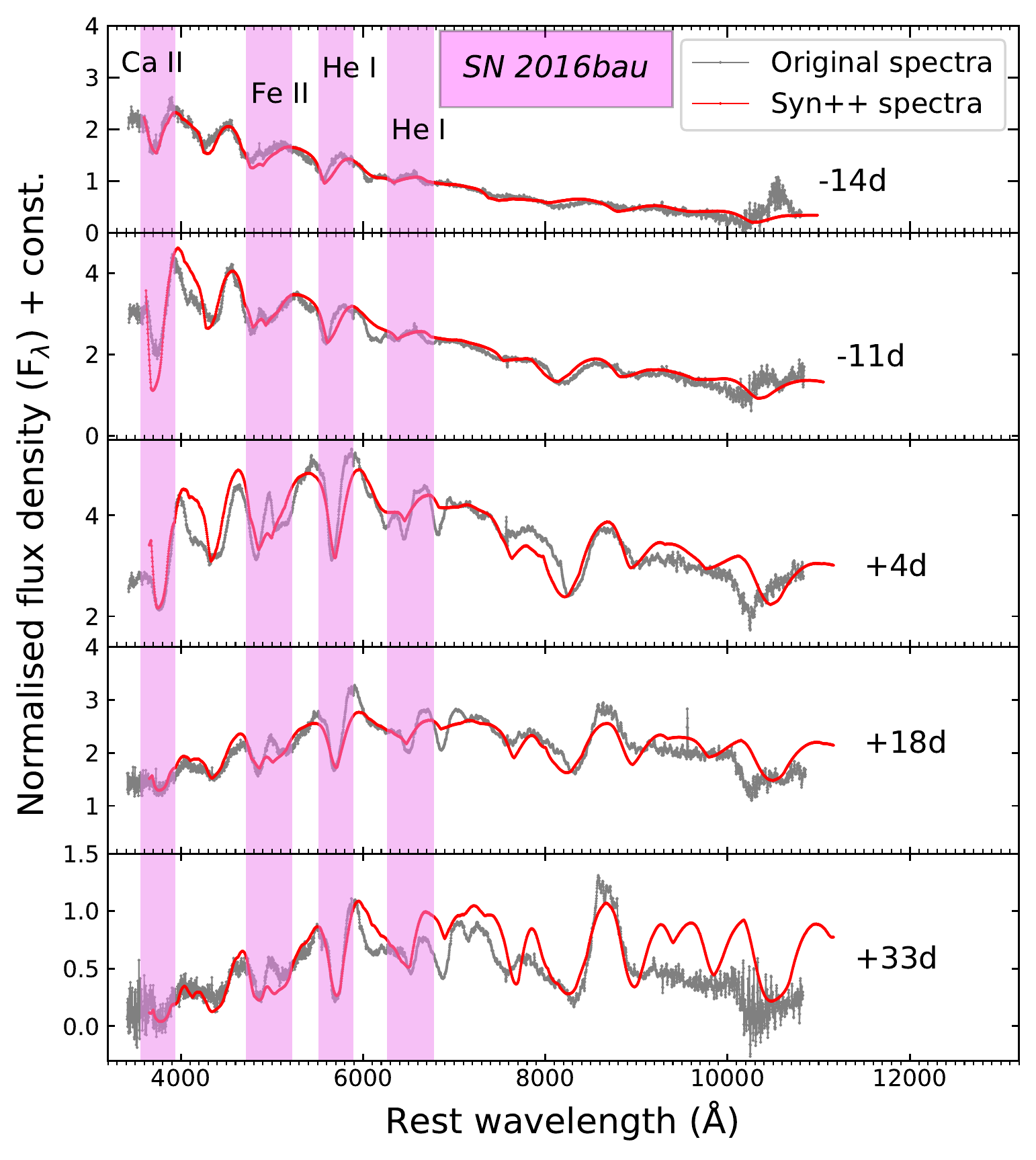}
    \caption{{\tt SYN++} modelling of the spectra of SN~2016bau at epochs $-14$\,d, $-11$\,d, $+4$\,d, $+18$\,d, and $+33$\,d.}
    \label{fig:syn_SN2016bau}
\end{figure}

After confidently identifying various spectral features, we tried to model the spectra of SN~2016bau at different epochs using {\tt SYN++}. The top panel of Figure~\ref{fig:syn_SN2016bau} shows the early-phase ($-14$\,d) spectrum of SN~2016bau. It contains many broad P~Cygni profiles of He~I, Ca~II~H\&K, the Ca~II~NIR triplet, and also some blended features of Fe~II. In this figure we have also presented the best-matching synthetic spectrum generated by {\tt SYN++}. The modelled spectrum easily reproduces the absorption features of Ca~II~H\&K, Ca~II~NIR, He~I, and the Fe~II multiplet. The photospheric velocity and blackbody temperature associated with the best-fit spectrum are 16,000\,km\,s$^{-1}$ and 9000\,K respectively. We performed {\tt SYN++} matching for four additional spectra which covers a period of $-11$\,d to +33\,d (subsequent panels of Fig.~\ref{fig:syn_SN2016bau}). With the passage of time, the SN expands and cools gradually, and its expansion velocity decreases slowly, so we see a gradual decrease in the fit parameters such as velocity and temperature in the later phases. The photospheric velocity during the phase of $-14$\,d to +33\,d varies from 16,000\,km\,s$^{-1}$ to 8000\,km\,s$^{-1}$ and the blackbody temperature ranges from 9000\,K to 4000\,K, in good agreement with values obtained photometrically from blackbody fits.

We also try to match the +121\,d spectrum of SN~2016bau with the 12\,$M_{\odot}$, 13\,$M_{\odot}$, and 17\,$M_{\odot}$ model spectra at +100\,d  from \citet[][]{Jerkstrand2015}, scaled with a factor of exp$(-2 \times \Delta t /111.4)$ \citep[][]{Jerkstrand2015}, where $\Delta t$ = 21, is the time difference of the epoch of model spectrum and the epoch of observed spectrum. We can see that all three models over--predict the observed fluxes from 3000\,\si{\angstrom} to around 6000\,\si{\angstrom}, beyond which the 12\,$M_{\odot}$ model spectrum seems to best describe the observed spectrum. It could nicely explain the [Ca~II] emission near 7300\,\si{\angstrom} and the Ca~II~NIR feature near 8500\,\si{\angstrom}. The 13\,$M_{\odot}$ model spectrum also produces the [Ca~II] emission but fails to explain the observed fluxes near the Ca~II-NIR triplet. The 17\,$M_{\odot}$ model spectrum overpredicts the flux throughout the entire wavelength range and thus fails to explain the spectrum of SN~2016bau. Hence, based on our analysis, a slightly low-mass progenitor ($\leq 12$\,$M_{\odot}$) is expected.

\begin{figure}
\includegraphics[width=\columnwidth]{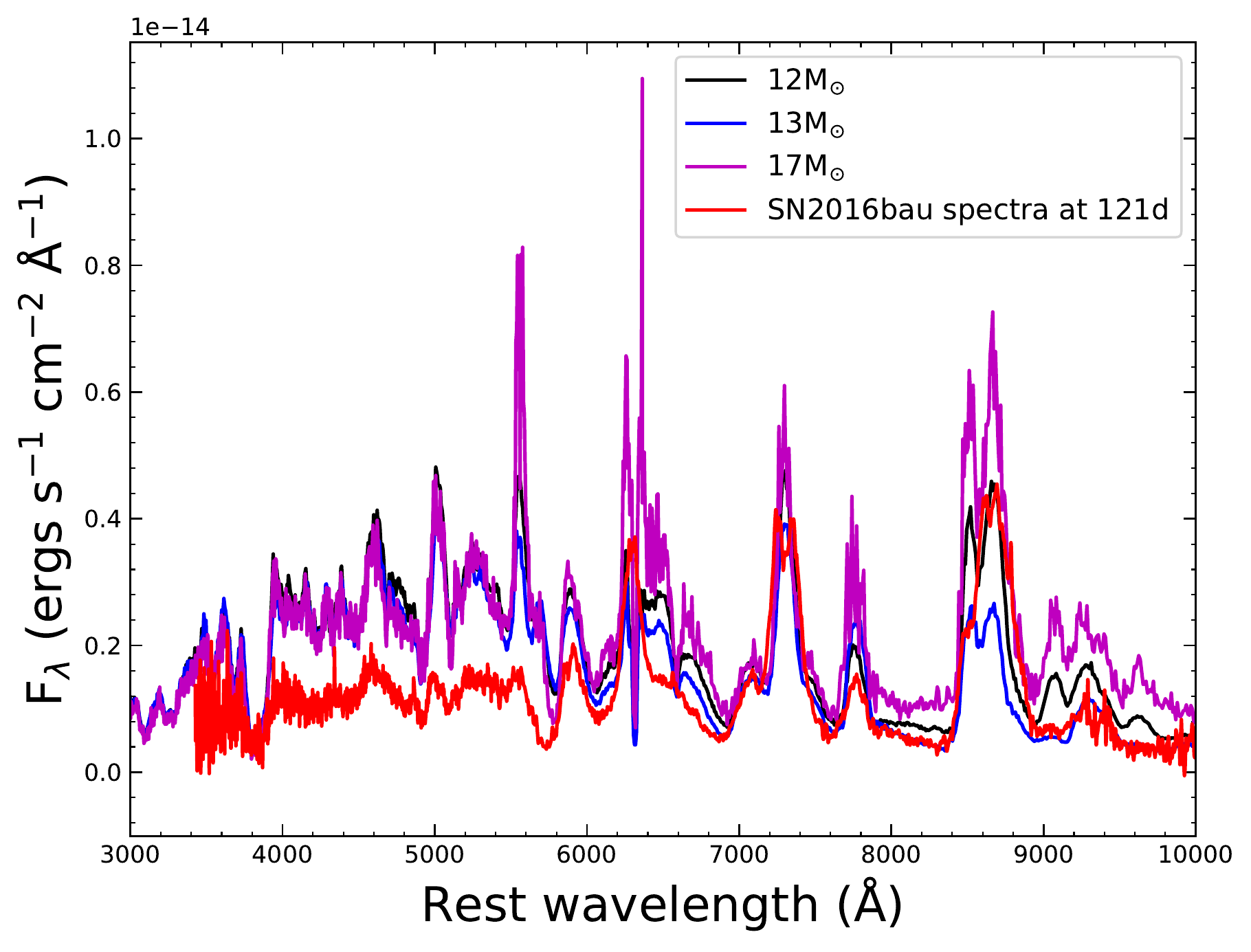}
\caption{The $t =  +121$\,d spectrum of SN~2016bau plotted along with the 12, 13, and 17\,$M_{\odot}$ models from \citet[][]{Jerkstrand2015} at 100\,d scaled with an exponential factor exp$(-2 \times 21/111.4)$. The 12\,$M_{\odot}$ model seems to best match the spectrum of SN~2016bau.}
\label{fig:jerkstrand_SN2016bau}
\end{figure}

\subsubsection{Velocity evolution of Various lines of SN~2016bau}
\label{subsec:Vel_evol_SN2016bau}

We used the blue-shifted absorption minima of P~Cygni profiles to obtain the velocities of He~I and Ca~II NIR lines. Figure~\ref{fig:vel_evolve}$b$ shows the He~I and Ca~II NIR velocity evolution of SN~2016bau. In the initial few days, the line velocities tend to decrease rapidly, but in later phases, the velocities decline gradually. At an early epoch of $-14$\,d, the velocities estimated using He~I 5876\,\si{\angstrom} and the Ca~II NIR triplet are $\sim$\,15,600\,km\,s$^{-1}$ and $\sim$\,11,400\,km\,s$^{-1}$, respectively. The velocities estimated using these two lines drop to $\sim$\,7800\,km\,s$^{-1}$ and $\sim$\,4400\,km\,s$^{-1}$ (respectively) at an epoch of +47\,d. Beyond +47\,d, the velocities continue to decline with a slower rate as compared to initial decline rate. 
Figure~\ref{fig:vel_evolve}$c,d$ show the comparison of these two line velocities with other well-studied SNe. Initially, the Ca~II NIR velocity declines very fast but in the later epochs, only gradual decline is seen. It evolves in a manner very similar to  iPTF13bvn, but in the late phases, the velocities are slower than other SNe. 
The velocity obtained using He~I 5876\,\si{\angstrom} of SN~2016bau evolves in a manner similar to that of other SNe~Ib, but much closer to SN~2009jf and iPTF13bvn. The He~I line velocity of SN~2016bau at +33\,d reaches $\sim$\,8200\,km\,s$^{-1}$, nearly equal to those of SN~2009jf and iPTF13bvn.

\section{MINIM modelling of the quasi-bolometric light curve}
\label{sec:lc_model}

In this section, we fit the radioactive decay (RD) and the magnetar (MAG)  powering mechanisms, as discussed by \citealt[][]{Chatzopoulous2013} (see also \citealt{Wheeler2017, Kumar2020, Kumar2021}), by employing the {\tt MINIM} \citep[][]{Chatzopoulous2013} code. {\tt MINIM} is a $\chi^{2}$-minimisation fitting code that utilises the Price algorithm \citep[][]{Brachetti1997}.  In the RD model, the radioactive decay of $^{56}$Ni and $^{56}$Co leads to the deposition of energetic gamma-rays that are assumed to thermalise in the homologously expanding SN ejecta and thus powering the light curve. In the MAG model, the light curves are powered by the energy released by the spin-down of a young magnetar, located in the centre of the SN ejecta. Following \citet[][]{Prentice2019}, we have adopted a constant opacity, $\kappa$ = 0.07\,cm$^2$\,g$^{-1}$, for both SN~2015ap and SN~2016bau.

\begin{table*}
\caption {Minimum $\chi^2$/dof parameters for SN~2015ap and SN~2016bau for the RD model.}
\label{tab:parameter_RD}
\begin{center}
{\scriptsize
\begin{tabular}{ccccccccccccc}
\hline\hline
	&	$M_{\mathrm{Ni}}$$^{a}$	&	$t_\mathrm{d}$$^{b}$	&	$A_{\mathrm{\gamma}}$$^{c}$	&	$M_{\mathrm{ej}}$$^{b}$ 	&	$\chi^2/\mathrm{dof}$	\\
	&	($M_{\odot}$)	&	(days)	&	&	($M_{\odot}$) 		\\
\hline
\hline
SN~2015ap\\
\hline
 whole light curve    &	0.094   $\pm$ 0.004	&	8.0  $\pm$  2.0	&	30.05   $\pm$   1.05 & 0.64 $\pm$ 0.3 & 6.2		\\
 +20\,d data    &	0.181   $\pm$ 0.006	&	14.5  $\pm$  0.3	&	5.3   $\pm$   0.3 & 2.1 $\pm$ 0.09 & 1.2		\\
\hline
SN~2016bau\\
\hline
 whole light curve    &	0.08   $\pm$ 0.01	&	20.5  $\pm$  0.7	&	7.8   $\pm$   0.5 & 2.8 $\pm$ 0.2 & 22.7		\\
  +20\,d data    &	0.065   $\pm$ 0.001	&	18.09  $\pm$  0.3	&	9.95   $\pm$   0.4 & 1.81 $\pm$ 0.07 & 1.2		\\

\hline\hline
\end{tabular}}
\end{center}

{$a$, mass of $^{56}$Ni synthesised;
$b$, effective diffusion timescale,}\\
{$c$, optical depth for the $\gamma$-rays measured 10\,d after the explosion;}\\
{$d$, ejecta mass, with $\kappa = 0.07$\,cm$^2$\,g$^{-1}$.}\\

\end{table*}

\begin{table*}
\caption{Minimum $\chi^2$/dof parameters for SN~2015ap and SN~2016bau for the MAG model.}
\label{tab:parameter_MAG}
\begin{center}
{\scriptsize
\begin{tabular}{ccccccccccccc}
\hline\hline
	&	$R_0$$^{a}$	&	$E_p$$^{b}$	  & 	$t_d$$^{c}$	&	$t_p$$^{d}$	 & $v_{\rm exp}$$^{e}$      & $M_{\rm ej}$$^{f}$  	& $P_i$$^{g}$   &    $B$$^{h}$    &   $\chi^2/\mathrm{dof}$\\
	&($10^{13}$\,cm) & ($10^{51}$\,erg) & (days)          & (days)         & ($10^3$\,km\,s$^{-1}$) & ($M_\odot$)    & (ms)          & ($10^{14}$\,G)   &		\\
\hline
\hline
SN~2015ap\\
\hline
      & 0.594   $\pm$  0.003 & 0.01210 $\pm$  0.00007 & 8.7  $\pm$  0.1 & 12.06   $\pm$  0.08 & 6.03  $\pm$  0.02  & 0.75$\pm$ 0.02 & 40.6 $\pm$0.1 & 25.5$\pm$ 0.2 & 1.54 \\
\hline
SN~2016bau\\
\hline
      & 7.9   $\pm$  0.7 & 0.00403 $\pm$  0.00007 & 8.5   $\pm$  0.1 & 15.1  $\pm$  0.1  & 6.2  $\pm$  0.5  & 1.26$\pm$ 0.02 & 70.4$\pm$ 0.6 & 52.6$\pm$ 0.8 & 1.4 \\
\hline\hline
\end{tabular}}
\end{center}

{$a$, progenitor radius; $b$, magnetar rotational energy; $c$, effective diffusion timescale (in days); $d$, magnetar spin-down timescale;}\\
{$e$, SN expansion velocity; $f$, $\kappa = 0.07$\,cm$^2$\,g$^{-1}$ is used; $g$, initial period of the magnetar; $h$, magnetic field of the magnetar.}\\
\end{table*}

\begin{figure}
	\includegraphics[width=\columnwidth]{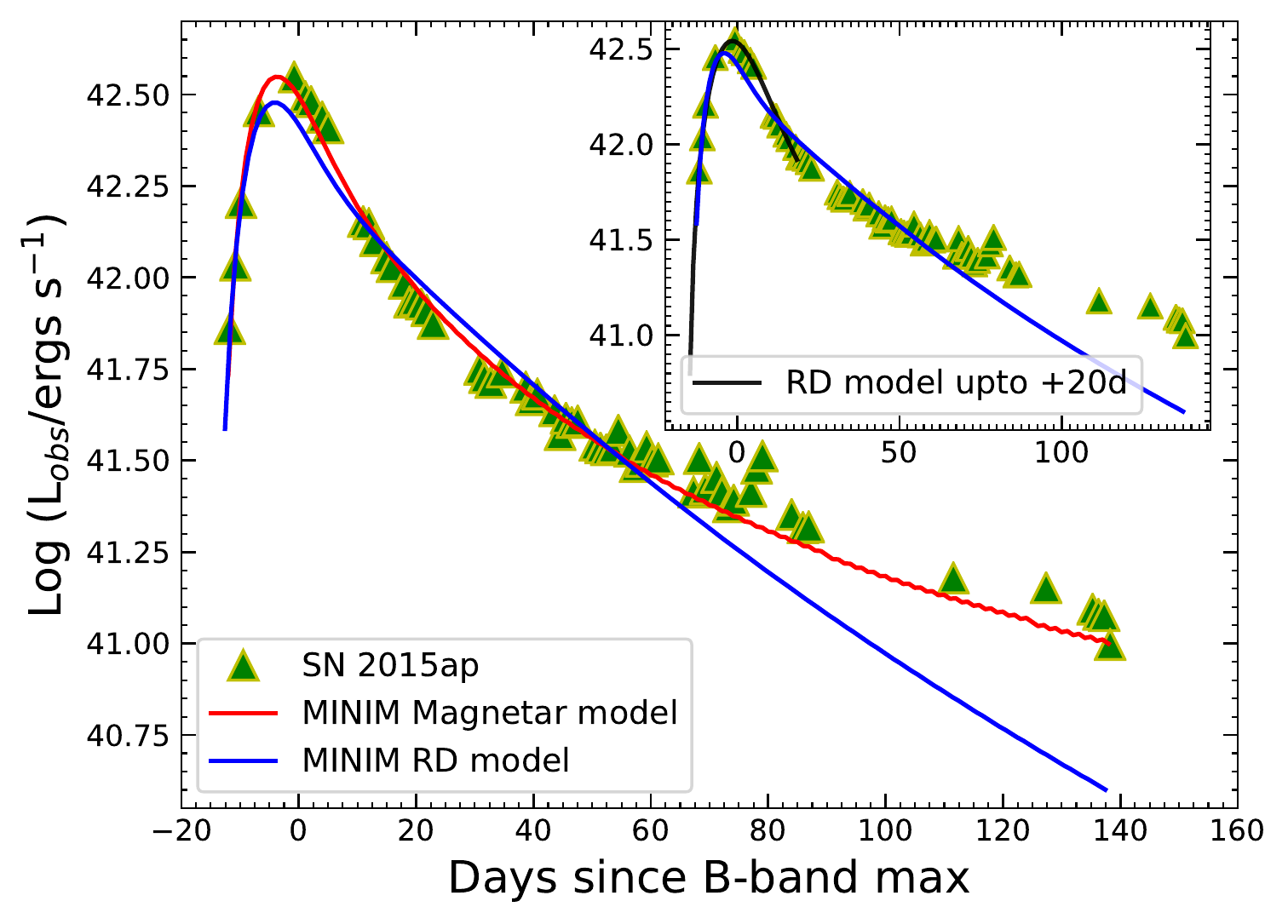}
 	\caption{{\tt MINIM} modelling of the quasi-bolometric light curve of SN~2015ap. The inset shows the RD model fitting only for the early phase (up to +20\,d).}
   	\label{fig:minim_SN2015ap}
\end{figure}

\begin{figure}
	\includegraphics[width=\columnwidth]{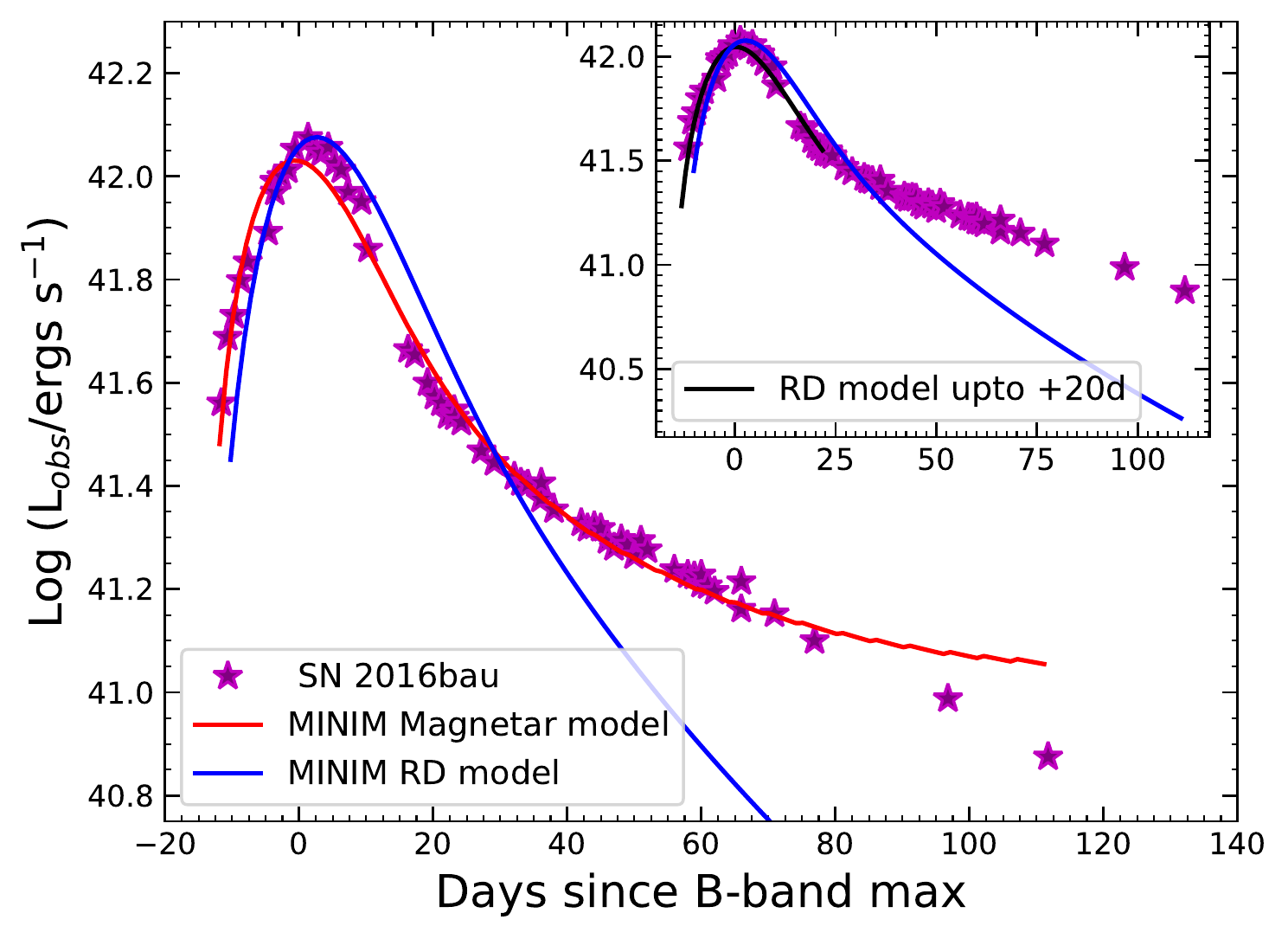}
 	\caption{{\tt MINIM} modelling of the quasi-bolometric light curve of SN~2016bau. The figure in the inset shows the RD model fitting only for the early phase (up to +20\,d).}
   	\label{fig:minim_SN2016bau}
\end{figure}

\subsection{SN~2015ap}

Figure~\ref{fig:minim_SN2015ap} shows the results of RD and MAG model fittings to the quasi-bolometric light curve of SN~2015ap. All of the fitted and calculated parameters are listed in Tables \ref{tab:parameter_RD} and \ref{tab:parameter_MAG}. The ejecta mass ($M_{\rm ej}$) in the RD and MAG models was calculated using Equation 1 from \citet[][]{Wheeler2015}. Although the RD model seems to reasonably fit the observed quasi-bolometric light curve, it seems unable to reproduce not only the observed peak luminosity,  but also the light curve at late phases, after +50\,d. The nickel mass and ejecta mass obtained from these models are somewhat smaller than the ones inferred directly from the observed rise time and peak luminosity in Sec.\ref{tempradvel_SN2015ap}. As the RD model is the prominent powering mechanism for normal SNe~Ib, we tried to fit the RD model only up to a relatively early phase ($\sim +20$\,d). We see in the inset of Figure~\ref{fig:minim_SN2015ap} that the early phase is nicely fitted in this case. A nickel mass ($M_{\rm Ni}$) of $0.181 \pm 0.006\,M_\odot$ and an ejecta mass ($M_{\rm ej}$) of $2.1 \pm 0.09\,M_\odot$ obtained through this fitting are close to the observed values listed in Sec.\ref{tempradvel_SN2015ap}. The fitted and calculated values from the photospheric phase are collected in Table~\ref{tab:parameter_RD}. 

The MAG model fits the whole observed light curve, both around peak brightness as well as during the late phase, better than the RD model. However, the fitted parameters seem to be unphysical, particularly the very slow initial rotation ($P_{i} \approx 40$\,ms) and the very low initial rotational energy ($E_{p} \approx 10^{49}$\,erg). This is not surprising given that the magnetar model contains {\it two} timescales, one for the rising part and another for the declining part of the light curve \citep{Chatzopoulous2013}.
Thus, the possibility of SN~2015ap powered by spin-down of a magnetar is less likely.

\subsection{SN~2016bau}

Figure~\ref{fig:minim_SN2016bau} shows the results of the RD and MAG model fittings to the quasi-bolometric light curve of SN~2016bau. The fitted and calculated parameters are listed in Tables \ref{tab:parameter_RD} and \ref{tab:parameter_MAG}. The RD model can fit the observed peak luminosity, but huge deviations from the observed light curve are seen in the later phases. The model also fails to match the observed stretch factor of the light curve. Like in the case of SN~2015ap, we also tried to fit only the early part, before +20\,d post-peak (see the inset in Fig.~\ref{fig:minim_SN2016bau}). 
In this case, the fitted parameters, such as the nickel mass of $0.065 \pm 0.002\,M_\odot$ and the ejecta mass of $1.81 \pm 0.07\,M_\odot$ (see Table~\ref{tab:parameter_RD}), are very close to the observed values in Section~\ref{tempradvel_SN2016bau}. Similar to SN~2015ap, the slow rotation ($P_i \approx 70$\,ms), low magnetar rotational energy ($E_p \approx 10^{48}$\,erg), and very high progenitor radius ($\sim 1200\,R_{\odot}$) make the MAG model physically unrealistic for SN~2016bau, despite the better fit to the whole light curve in Figure~\ref{fig:minim_SN2016bau}.

\subsection{Summary of {\tt MINIM} modelling}
The failure of the semi-analytical RD models to simultaneously fit the early and late parts of the light curves of SN~2015ap and SN~2016bau highlights the issues related to the validity of such kinds of models in the case of stripped-envelope SNe. The difficulty in explaining the late-phase decline rate with the assumed diffusion model have already been explored by \citet{Wheeler2015}. 

Figures~\ref{fig:minim_SN2015ap} and \ref{fig:minim_SN2016bau} reveal another issue: the model light curve is steeper after +30--40\,d than the observed one. This indicates that the early- and late-phase data cannot be fitted simultaneously with the same model parameters. 
Even though the nickel mass of the second model fits (insets in Figs.~\ref{fig:minim_SN2015ap} and \ref{fig:minim_SN2016bau}) the observed peak luminosity, the decline rate, which is related to the ejecta mass (see Eq. 1 of \cite{Wheeler2015}) is clearly too fast, suggesting an underestimated $M_{\rm ej}$. 

\begin{figure}
	\includegraphics[width=\columnwidth]{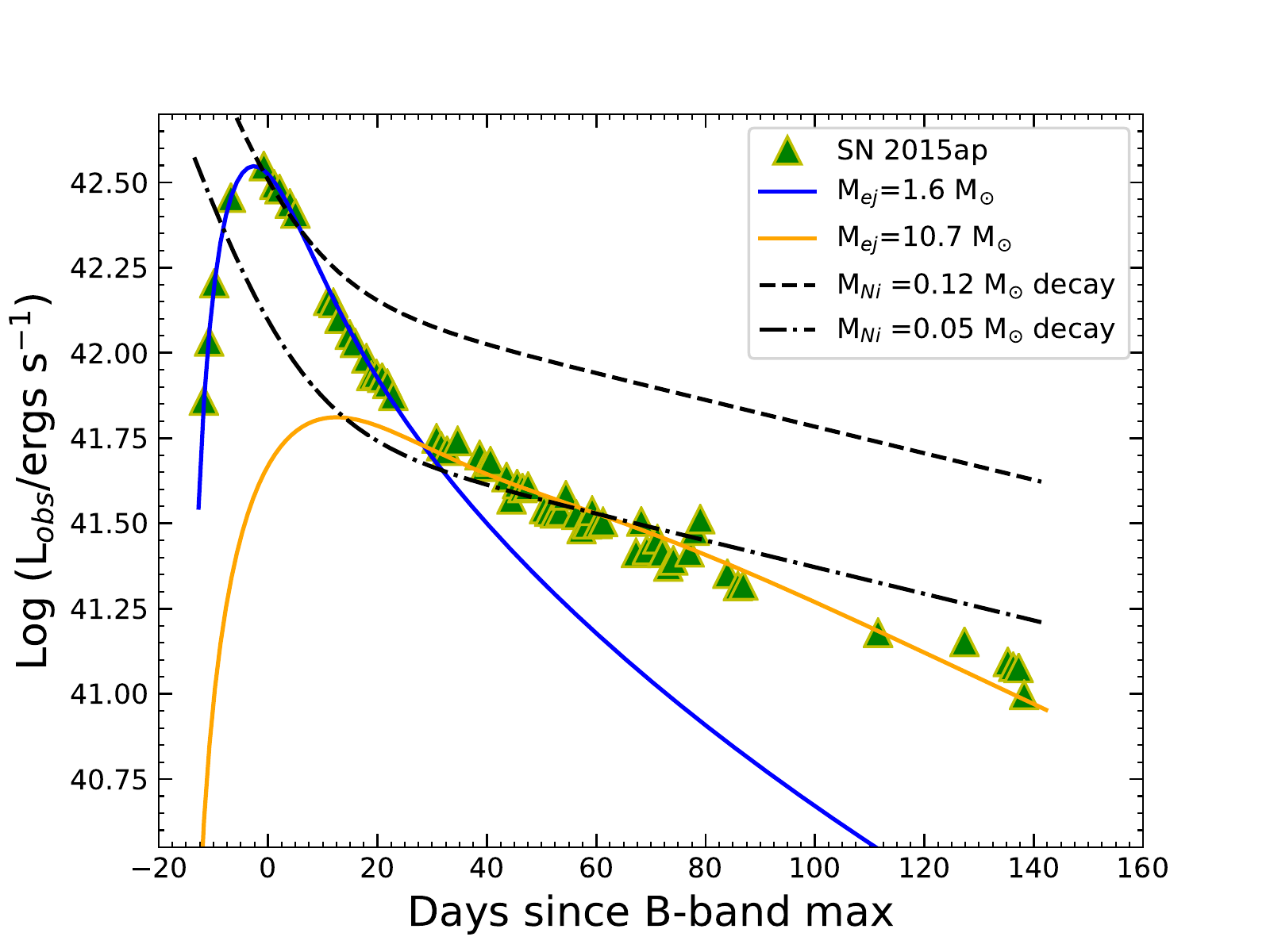}
 	\caption{Detailed {\tt MINIM} modelling to the quasi-bolometric light curve of SN~2015ap. All of the models with different $M_{\rm ej}$ and $M_{\rm Ni}$ are shown together. The models with $M_{\rm ej} = 1.6\,M_{\odot}$ and 10.7\,$M_{\odot}$ show the RD model fittings considering only the early- and late-phase data, respectively. The other two models show the effect of the amount of nickel present. }
   	\label{fig:minim_SN2015ap_tot}
\end{figure}

This issue is very probably related to the assumption of the constant ejecta density profile (and also the constant opacity). The less steep late part of the light curve needs more ejecta mass to trap the heating gamma-rays originating from the Ni and Co decay. The early part, however, suggests ejecta that dilute much faster than what can fit the late part. Within the context of the constant-density model, this dichotomy means that fitting only the early part results in a lower ejecta mass compared to fitting only the late part (Fig.\ref{fig:minim_SN2015ap_tot}). 

Since this issue cannot be solved self-consistently with the constant-density model, and it can plague the mass estimates that consider only the early or late part of the light curve, we investigate it further in the following sections by applying more realistic models for the progenitors and the SN light curves. 

\section{MESA modelling of a 12 $M_{\odot}$ ZAMS progenitor star}
\label{sec:mesa}

Adoption of a 12\,$M_{\odot}$ ZAMS progenitor star for SN~2015ap is primarily based on the results of comparison of the three models from \citet[][]{Jerkstrand2015} and also in the literature; in addition, it supports the observed amount of ejecta mass for SN~2015ap. The ejecta mass for SN~2016bau was calculated to be $\sim 1.6\,M_{\odot}$. On account of such an ejecta mass and the result of \citet[][]{Jerkstrand2015} spectral model matching, we also choose a 12\,$M_{\odot}$ ZAMS star as a possible progenitor for SN~2016bau.

We first evolve the 12\,$M_{\odot}$ ZAMS star until the onset of core-collapse, using the one-dimensional stellar evolution code {\tt MESA}, version 11701\citep[][] {Paxton2011,Paxton2013,Paxton2015,Paxton2018}. We do not consider rotation and assume an initial metallicity of $Z = 0.02$. Convection is modelled using the mixing theory of \citet[][]{Henyey1965}, adopting the Ledoux criterion. We set the mixing-length parameters to $\alpha = 3.0$ in the region where the mass fraction of hydrogen is greater than 0.5, and set it to 1.5 in the other regions. Semi-convection is modelled following \citet[][]{Langer1985} with an efficiency parameter of $\alpha_{\mathrm{sc}} = 0.01$. For the thermohaline mixing, we follow \citet[][]{Kippenhahn1980}, and set the efficiency parameter as $\alpha_{\mathrm{th}} = 2.0$. We model the convective overshooting with the diffusive approach of \citet[][]{Herwig2000}, with $f= 0.01$ and $f_0 = 0.004$ for all the convective core and shells. We use the \say{Dutch} scheme for the stellar wind, with a scaling factor of 1.0. The \say{Dutch} wind scheme in MESA combines results from several papers. Specifically, when $T_{\mathrm{eff}} > 10^4$\,K and the surface mass fraction of hydrogen is greater than 0.4, the results of \citet[][]{Vink2001} are used, and when $T_{\mathrm{eff}} > 10^4$\,K and the surface mass fraction of hydrogen is less than 0.4, the results of \citet[][]{Nugis2000} are used. In the case when $T_{\mathrm{eff}} < 10^4$\,K, the \citet[][]{dejager1988} wind scheme is used. 

SNe~Ib have been considered to originate from  massive stars which lose almost all of their hydrogen envelope, most probably due to binary interaction \citep[e.g.,][]{Yoon2010, Dessart2012, Eldridge2016, Ouchi2017}. Here, in order to produce such a stripped model, we artificially strip the hydrogen envelope, mimicking the binary interaction. Specifically, after evolving the model until the exhaustion of helium, we impose an artificial mass-loss rate of $\dot{M} \gtrsim 10^{-4}\,M_{\odot}\,\mathrm{yr}^{-1}$ until the total hydrogen mass of the star goes down to 0.01\,$M_{\odot}$. After the hydrogen mass reaches the specified limit, we switch off the artificial mass loss and evolve the model until the onset of core-collapse. At the time of core-collapse, our model has a total mass of $3.42\,M_{\odot}$.

\section{Explosions of modelled progenitors using SNEC and STELLA}
\label{sec:snec}

In this section we briefly discuss the assumptions and setups to produce  artificial explosions using {\tt SNEC} \citep[][]{Morozova2015} and {\tt STELLA} \citep[][]{Blinnikov1998, Blinnikov2000, Blinnikov2006} for SN~2015ap and SN~2016bau.

\subsection{SN~2015ap}
Using the progenitor model on the verge of core-collapse obtained through {\tt MESA}, we then carried out the radiation hydrodynamical simulations. For this purpose, we use the publicly available codes {\tt SNEC} and {\tt STELLA}.

{\tt SNEC} is a one-dimensional Lagrangian hydrodynamic code, which also solves radiation energy transport with the flux-limited diffusion approximation. The code generates the bolometric light curve and the photospheric velocity evolution of the SN, along with other observed parameters.
The setup for the calculation using {\tt SNEC} closely follows \citet[][]{Ouchi2019}. Here, we briefly summarise the important parameters and modifications made to \citet[][]{Ouchi2019}. First, we excise the innermost $1.4\,M_{\odot}$ before the explosion, assuming that it collapses to form a neutron star. The number of cells is set to be 70. Although this number is relatively small, we have confirmed that the light curve and photospheric velocity of the SN are well converged in the time domain of interest.
We tried following two possible powering mechanisms for SN~2015ap using {\tt SNEC}, as follows.

\begin{figure}
\includegraphics[width=\columnwidth]{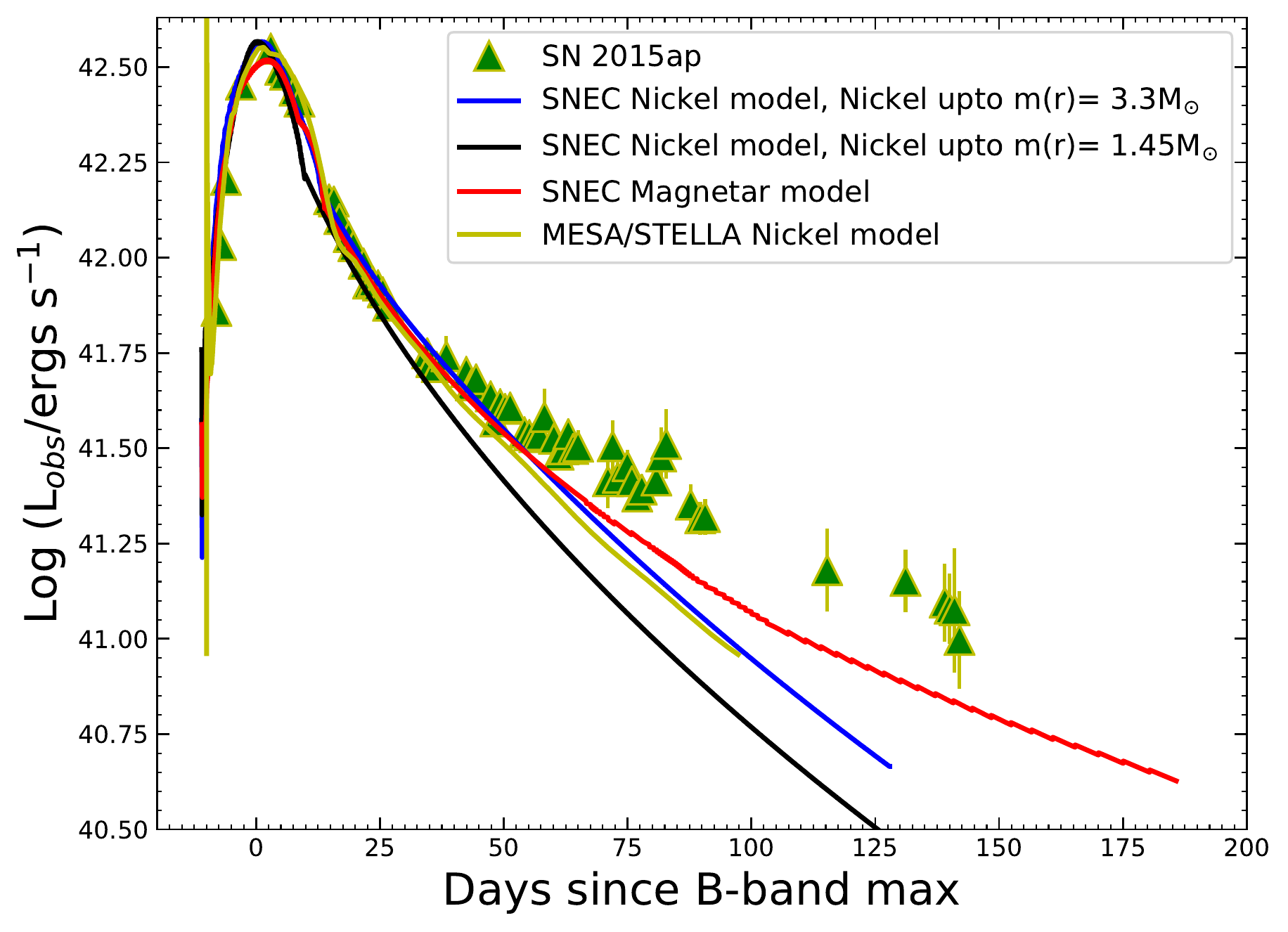}
\caption{Comparison of quasi-bolometric light curve of SN~2015ap with those obtained using {\tt SNEC} by taking into account the Ni and Co decay model and the magnetar model. This figure also depicts the result of the Ni--Co decay model obtained using {\tt STELLA}.}
\label{fig:lum}
\end{figure}

\begin{figure}
\includegraphics[width=\columnwidth]{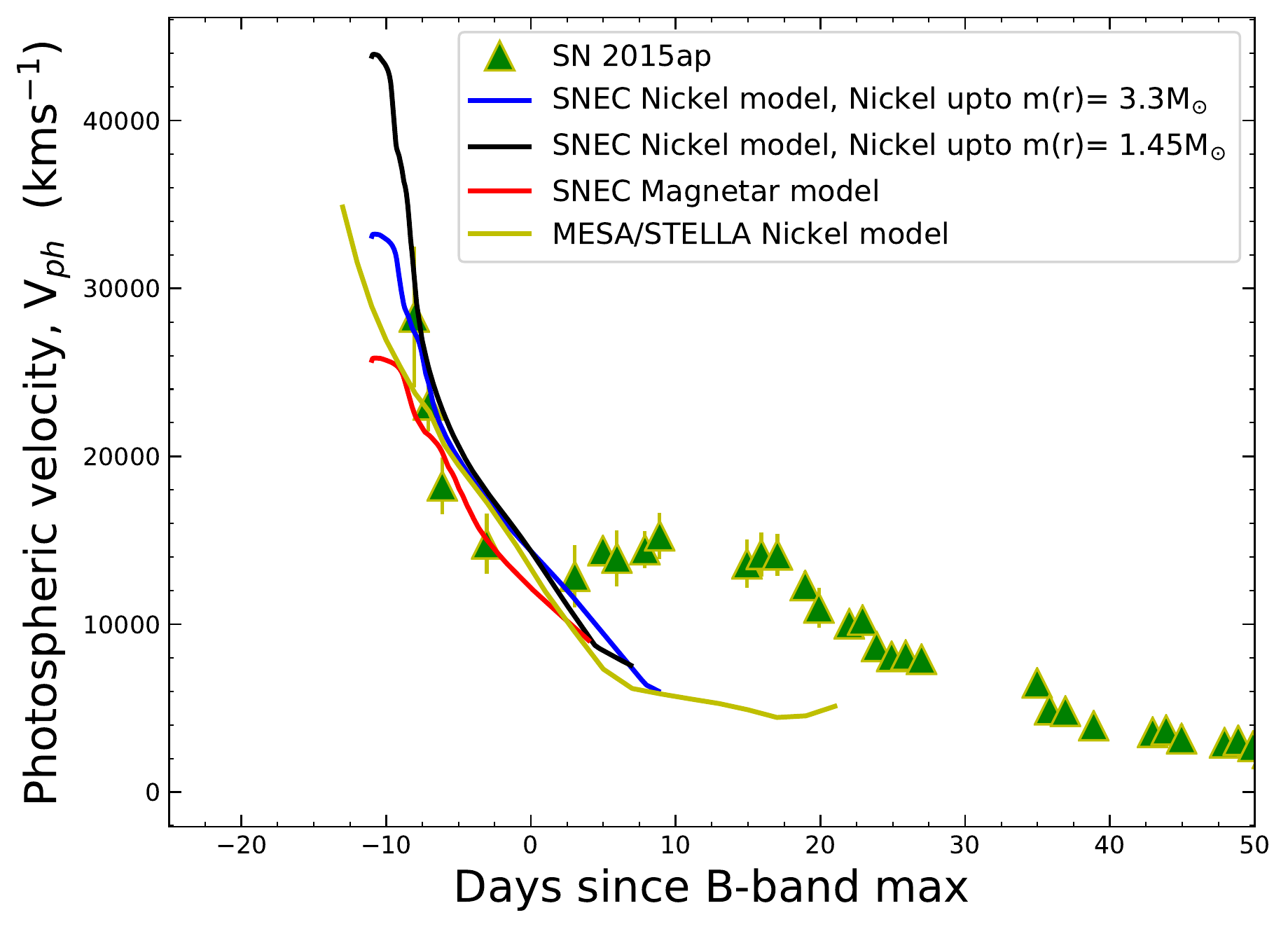}
\caption{Comparison of the observed velocity evolution of SN~2015ap with that produced by {\tt SNEC} using two models. We also show a comparison of the observed velocity evolution produced with {\tt STELLA}.}
\label{fig:velocity}
\end{figure}

\subsubsection{Ni--Co Decay}

The radioactive decay of Ni and Co is considered to be the most prominent mechanism for the powering of light curves of SNe~Ib \citep[e.g.,][]{Karamehmetoglu2017}. {\tt SNEC} incorporates this model by default. Here, we provide the setup of the explosion parameters to incorporate the Ni--Co decay model. 
The code does not include a nuclear-reaction network, and $^{56}$Ni is given by hand. We considered two scenarios of nickel distribution. In one case, the mass of Ni is set to be $M_{\mathrm{Ni}}=0.135\,M_{\odot}$ and distributed from the inner boundary up to the mass coordinate $m (r) = 3.3\,M_{\odot}$. For this model the explosion is simulated as a {\tt piston}, with the first two computational cells of the profile boosted outward with a velocity of $4.0 \times 10^{9}$\,cm\,s$^{-1}$ for a time interval of 0.01\,s. The total energy ($E_{\rm tot}$) of the model is $3.7 \times 10^{51}$\,erg.  For the second case, the mass of Ni is set to be $M_{\mathrm{Ni}}=0.1\,M_{\odot}$ and distributed near the centre, from the inner boundary up to the mass coordinate of $m (r) = 1.45\,M_{\odot}$. The explosion is simulated as a {\tt piston}, with the first two computational cells of the profile boosted outward with a velocity of $5.0 \times 10^{9}$\,cm\,s$^{-1}$ for a time interval of 0.01\,s. The total energy of the model in this case is $6.5 \times 10^{51}$\,erg. Thus, our results provide a range of $M_{\mathrm{Ni}}$ and total energy, depending on the distribution of nickel mass.

\subsubsection{Spin-down of a magnetar} 
We also tried a magnetar-powering mechanism for the light curve as SN~2015ap; it shows some resemblance to SN~2008D, which has broader features in its early-time spectra and also some X-ray emission in the later phases. A few SNe~Ib are also explained by magnetar models, one such example being SN~2005bf \citep[][]{Maeda2007}. For this model, the explosion is the {\tt piston} type, with the first two computational cells of the profile boosted outward with a velocity of $3.3 \times 10^{9}$\,cm\,s$^{-1}$ for a time interval of 0.01\,s. The total energy of the model in this case is $2.2 \times 10^{51}$\,erg. The most important change made to \citet[][]{Ouchi2019} is that we add the magnetar heat to the ejecta.

Following \citet[][]{Metzger2015}, the magnetar spin-down luminosity is given by
\[
L_{\mathrm{sd}} = L_{\mathrm{sd_i}} (1 + t/t_{\mathrm{sd}})^{-2}.
\]
\noindent
Here, $L_{\mathrm{sd_i}}$ is the spin-down luminosity at $t=0$, and $t_{\mathrm{sd}}$ is the initial spin-down time. We inject this luminosity into the whole ejecta above the mass cut uniformly in mass.
For the initial spin-down luminosity, we assume $L_{\mathrm{sd{_{i}}}} = 1.3 \times 10^{43}$\,erg\,s$^{-1}$, while for the initial spin-down time, we assume $t_{\mathrm{sd}} = 12$\,d. In this model, we do not include the effect of Ni heating. Following \citet[][]{Metzger2015} (Equations 2 and 3), corresponding to $L_{\mathrm{sd{_{i}}}} = 1.3 \times 10^{43}$\,erg\,s$^{-1}$ and $t_{\mathrm{sd}} = 12$\,d, we obtain a magnetic field ($B$) of $5.1 \times 10 ^{14}$\,G and an initial period ($P_i$) of 43.06\,ms for the modelled magnetar. These values of $B$ and $P_i$ are very close to those obtained using {\tt MINIM}.

We also use the public version of {\tt STELLA}, available with {\tt MESA}. The default radioactive decay of $^{56}$Ni and $^{56}$Co as a powering mechanism is used for SN~2015ap. Nearly similar parameters as in the case of {\tt SNEC} are used for {\tt STELLA}. The {\tt MESA} setups are unchanged for {\tt STELLA} calculations. We use a total energy after explosion of $3.6 \times 10^{51}$\,erg and a $^{56}$Ni mass of $0.193\,M_\odot$, which is slightly more than the $^{56}$Ni mass used in {\tt SNEC}. In order to avoid the numerical problem caused by the high-velocity material, we removed the outer layer of the progenitor where the density is less than $10^{-5}$\,g\,cm$^{-3}$ \citep[see also][]{Moriya2020}.

\subsubsection{Summary of the modelling}

Figure~\ref{fig:lum} shows the comparison of the observed quasi-bolometric luminosity with that produced by {\tt SNEC}. We find that the radioactive decay models with $^{56}$Ni mass in the range 0.1--0.135\,$M_\odot$ and total energy in the range (3.7--6.5) $\times 10^{51}$\,erg could nicely explain the light curve. Our results also signify that the distribution of $^{56}$Ni mass plays an important role for explaining the observed light curves.
The magnetar model could also explain the quasi-bolometric light curve, but we do not see very strong evidence that SN~2015ap is powered by a magnetar. This figure also shows the results of {\tt STELLA} calculations. We see that, although the explosion energy is similar, we need a slightly higher amount of nickel to properly match the observed light curve of SN~2015ap.
  
Figure~\ref{fig:velocity} illustrates a comparison of observed photospheric velocity evolution produced by blackbody fitting with that produced by {\tt SNEC}. We see that the Ni--Co decay and magnetar models initially show higher velocities, but in the later epochs they well replicate the observed photospheric velocities.  This figure also shows the Fe~II 5169\,\si{\angstrom} velocity evolution produced by {\tt STELLA}. Owing to the unambiguous absence of Fe~II 5169\,\si{\angstrom} features in SN~2015ap, we used the photospheric velocity obtained through a blackbody fit for our comparison of Fe~II 5169\,\si{\angstrom} line velocities. We can see a very good match between the {\tt STELLA} velocities and observed ones. The modelling parameters, along with the observed values, are listed in Table \ref{tab:parameter_model}.

\subsection{SN~2016bau}
Using a similar progenitor model on the verge of core-collapse, obtained through {\tt MESA}, as in the case of SN~2015ap, we carried out the radiation hydrodynamical simulations using {\tt SNEC} and {\tt STELLA}. For calculations using {\tt SNEC}, the setup closely follows that of \citet[][]{Ouchi2019}.  First, we excise the innermost $1.53\,M_{\odot}$ before the explosion, assuming that it collapses to form a neutron star. The number of cells is set to 70. We try two possible powering mechanisms for SN~2016bau using {\tt SNEC}, as follows.

\begin{table*}
\caption{Observed and modelled parameters for SN~2015ap and SN~2016bau}
\label{tab:parameter_model}
\begin{center}
{\scriptsize
\begin{tabular}{ccccccccccccc}
\hline\hline
	&	$M_{\mathrm{Ni}}$	&	$M_{\mathrm{ej}}$	&	$E_{\mathrm{tot}}$	\\
	&	($M_{\odot}$)	&	($M_{\odot}$)	& $10^{51}$\,erg		\\
\hline
\hline
SN~2015ap\\
\hline
 Arnett's model$^{a}$    &	0.14   $\pm$ 0.02	&	2.2  $\pm$ 0.6	&	* \\

 {\tt SNEC} (Ni distributed up to $M(r) = 3.3\,M_{\odot}$)    &	    0.135	&	  2.02	&	   3.7 	\\
 
 {\tt SNEC} (Ni distributed up to $M(r) = 1.45\,M_{\odot}$)    &	    0.1	&	  2.02	&	   6.5 	\\

 {\tt SNEC}  Magnetar model   &	0.0	&	  2.02	&	   2.2 	\\

 From {\tt STELLA}    &	0.193	& 1.92		&	3.6	\\
\hline
SN~2016bau\\
\hline
 Arnett's model$^{a}$   &	0.055   $\pm$ 0.006	&	1.6  $\pm$ 0.3	&	** \\


{\tt SNEC} (Ni distributed up to $M(r) = 3.3\,M_{\odot}$)    &	    0.045	&	  1.89	&	   1.23 	\\
 
{\tt SNEC} (Ni distributed up to $M(r) = 1.57\,M_{\odot}$)    &	    0.03	&	  1.89	&	   1.93 	\\

{\tt SNEC}  Magnetar model   &	0.0	&	1.89	&	0.8 	\\

 From {\tt STELLA}    &	0.065  	&	1.89  	&	1.6	\\
\hline\hline
\end{tabular}}
\end{center}
{$a$: Calculated using $t_{\rm rise}$, $\kappa = 0.07$\,cm$^2$\,g$^{-1}$, and $L_{\rm peak}$.}\\
{$*$: Instead, a kinetic energy of the ejecta $(E_{k}) = 1.05 \times 10^{51}$\,erg is obtained.}\\
{$**$: Instead, a kinetic energy of the ejecta $(E_{k}) = 0.24 \times 10^{51}$\,erg is obtained.}\\
\end{table*} 

\subsubsection{Ni--Co Decay}
The setup for the calculation using {\tt SNEC} is similar to that of SN~2015ap.
Considering the radioactive decay of Ni--Co the most prominent mechanism for powering light curves of SNe~Ib, we employed this model as the powering mechanism for SN~2016bau. Here we briefly describe the setup of the explosion parameters incorporated in the Ni--Co decay model.

We considered two cases of $^{56}$Ni mass distribution. In the first case, the mass of $^{56}$Ni synthesised is set to be $M_{\mathrm{Ni}}=0.045\,M_{\odot}$. Then, it is distributed from the inner boundary up to the mass coordinate of $M(r) = 3.3\,M_{\odot}$. Thereafter, we simulate the explosion as a {\tt piston}, with the first two computational cells of the profile boosted outward with a velocity of $4.2 \times 10^{9}$\,cm\,s$^{-1}$ for a time interval of 0.01\,s. The model has a total energy of $1.3 \times 10^{51}$\,erg. For the second case, the mass of $^{56}$Ni synthesised is set to be $M_{\mathrm{Ni}}=0.03\,M_{\odot}$ and distributed from the inner boundary up to the mass coordinate of $M(r) = 1.57\,M_{\odot}$. Thereafter, we simulate the explosion as a {\tt piston}, with the first two computational cells of the profile boosted outward with a velocity of $4.9 \times 10^{9}$\,cm\,s$^{-1}$ for a time interval of 0.01\,s, and the model has a total energy of $1.93 \times 10^{51}$\,erg. We find that, depending on the $M_{\mathrm{Ni}}$ distribution, the radioactive decay models with $^{56}$Ni mass in the range 0.03--$0.045\,M_\odot$ and total energy in the range (1.23--1.93) $\times 10^{51}$\,erg can nicely explain the light curve.

\begin{figure}
\includegraphics[width=\columnwidth]{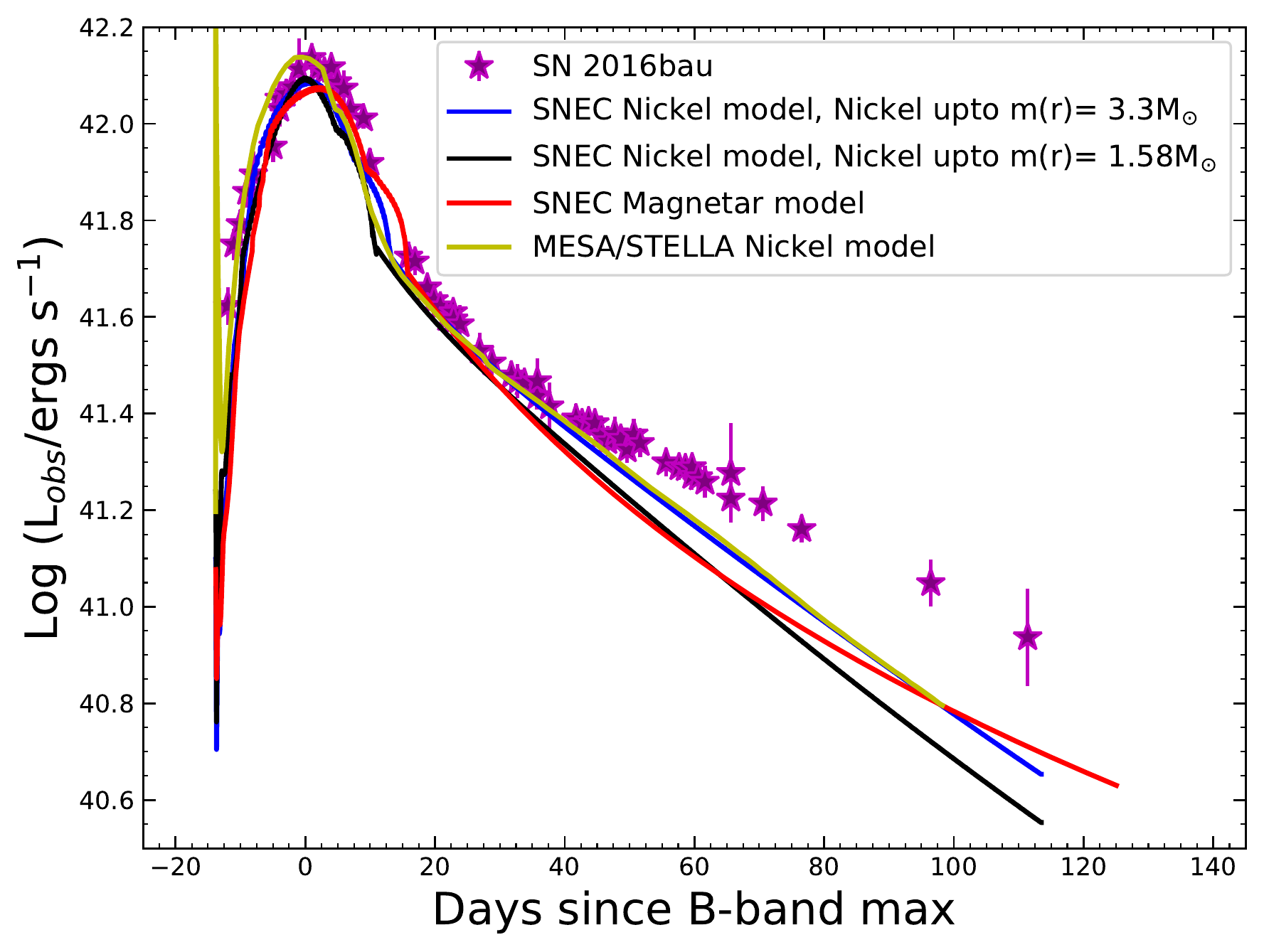}
\caption{Comparison of the quasi-bolometric light curve of SN~2016bau and light curves produced by {\tt SNEC} and {\tt STELLA}, considering the radioactive decay model.}
\label{fig:lum_SN2016bau}
\end{figure}

\begin{figure}
\includegraphics[width=\columnwidth]{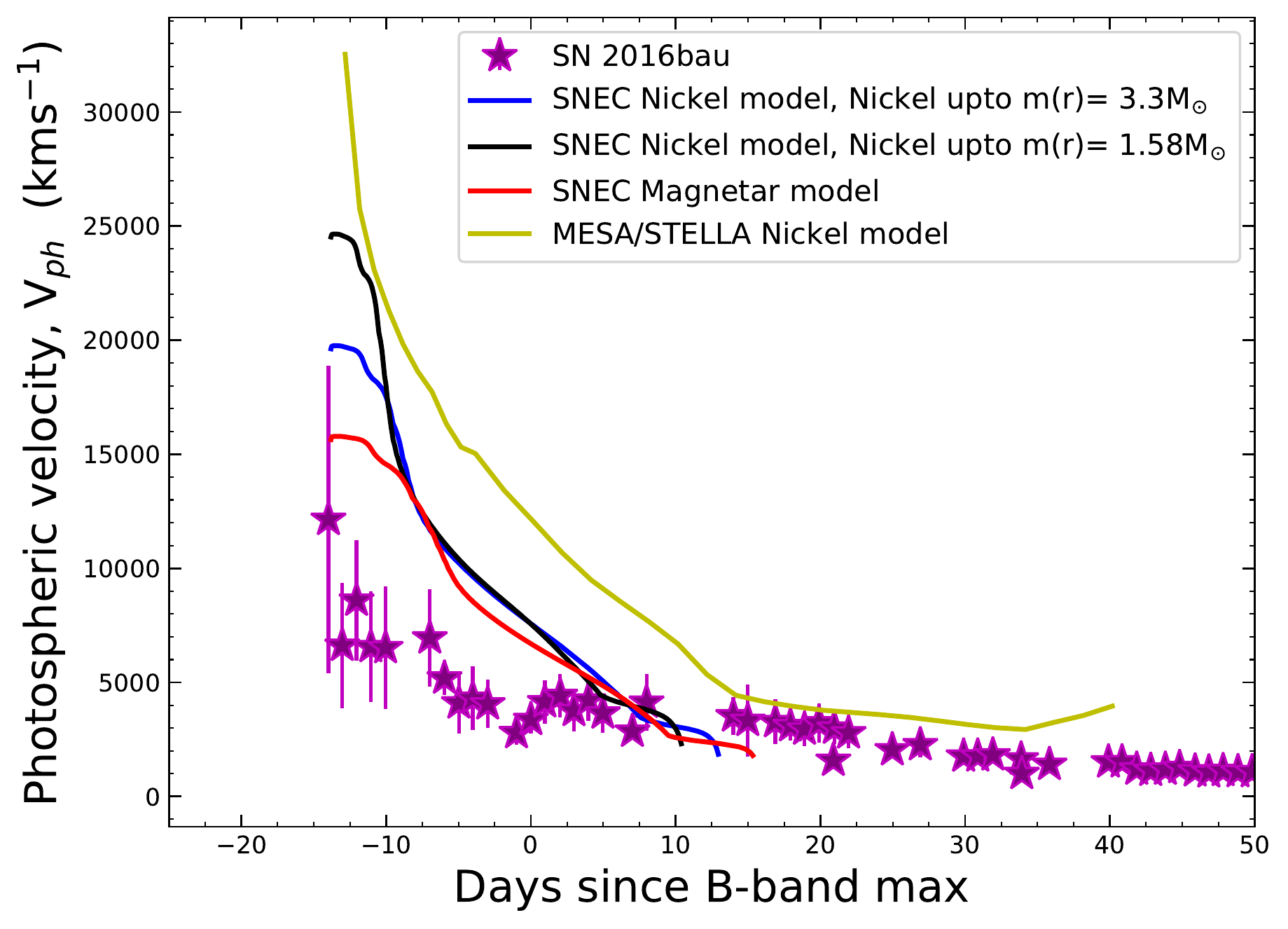}
\caption{Comparison between the observed velocity evolution of SN~2016bau and the photometric velocity evolution produced by {\tt SNEC} and {\tt STELLA}, considering the radioactive decay model.}
\label{fig:velocity_SN2016bau}
\end{figure}

\subsubsection{Spin-down of a magnetar} 
We also tried the magnetar powering mechanism for SN~2016bau. The explosion is simulated as a {\tt piston}, with the first two computational cells of the profile boosted outward with a velocity  of $3.6 \times 10^{9}$\,cm\,s$^{-1}$ for a duration of 0.01\,s and a total model energy of $0.8 \times 10^{51}$\,erg. Like SN~2015ap, we inject $L_{\mathrm{sd_i}} = 0.4 \times 10^{43}$\,erg\,s$^{-1}$ to the ejecta above the mass cut uniformly in mass. We assume $t_{\mathrm{sd}} = 16.0$\,d and do not include the effect of Ni heating. Corresponding to $L_{\mathrm{sd{_{i}}}} = 0.4 \times 10^{43}$\,erg\,s$^{-1}$ and $t_{\mathrm{sd}} = 16$\,d, we obtain a magnetic field ($B$) of $7.9 \times 10^{14}$\,G and an initial period ($P_i$) of 71.8\,ms for the modelled magnetar. These values of $B$ and $P_i$ are very close to those obtained using {\tt MINIM}.

We also perform {\tt STELLA} calculations for SN~2016bau, by employing the default radioactive decay of $^{56}$Ni and $^{56}$Co powering mechanism. We used similar parameters as in the case of {\tt SNEC} and {\tt STELLA}. The {\tt MESA} setups are unchanged for {\tt STELLA} calculations. We used a total energy after explosion of $1.6 \times 10^{51}$\,erg and a $^{56}$Ni mass of $0.065\,M_\odot$, slightly above the $^{56}$Ni mass used in {\tt SNEC}.

\subsubsection{Summary of the modelling}

Figure~\ref{fig:lum_SN2016bau} shows the comparison of the observed quasi-bolometric luminosity with that produced by {\tt SNEC} for the radioactive decay and magnetar spin-down models. We find that the radioactive decay model from {\tt SNEC} could successfully explain the light curve. The magnetar model also shows a good match, but we do not see any significant signs of SN~2016bau being powered by a magnetar. This figure also shows the results of {\tt STELLA} calculations; they also match the observed light curve of SN~2016bau nicely, with parameters similar to those of {\tt SNEC}.

Figure~\ref{fig:velocity_SN2016bau} shows a comparison of the observed photospheric velocity evolution produced by blackbody fitting with that produced by the radioactive decay model and the magnetar spin-down model using {\tt SNEC}. Here, the models show high initial velocities which drop at later epochs. We see that the velocities produced by the models initially deviate from the observed photospheric velocities, but tend to follow velocities similar to the observed ones at later epochs. The figure also shows a comparison the Fe~II 5169\,\si{\angstrom} line velocity obtained using {\tt STELLA}  with the photospheric velocity obtained through blackbody fits. Similar to the case of SN~2015ap, this SN also lacks unambiguous features of the Fe~II 5169\,\si{\angstrom} line.

From Figure~\ref{fig:velocity_SN2016bau}, it is evident that our model overestimates the velocity by a factor of nearly two. The $M_{\rm ej}$ from Arnett's model is $\sim 1.6\,M_{\odot}$. The diffusion time is given by $(M^3/E)^{1/4}$. Since this combination is fixed, $M^3/E$ is roughly constant. But $M^3/E \approx M^3/(Mv^2) = M^2/v^2 \approx$ constant. Thus, we have $M \approx v$ to meet the observational constraints. So, if we need to decrease the velocity by a factor of two, the ejecta mass would also decrease by the same factor. Then the He star mass is likely $\sim 2$--2.5\,$M_{\odot}$, which is at the boundary between a SN and a non-SN. Since $E \approx Mv^2 \approx v^3$, the energy may go down quite substantially in this case.

\section{Discussion}
\label{sec:Discussions}
We present a detailed photometric and spectroscopic analysis of two Type Ib SNe, namely SN~2015ap and SN~2016bau. From our analysis, SN~2015ap is an intermediate-luminosity normal SN~Ib, while SN~2016bau is highly extinguished by host-galaxy dust. In this section, we discuss the major outcomes of our present analysis.

The photometric properties of both the SNe were analysed by determining the bolometric luminosity of their light curves. For SN~2015ap, we calculate a $^{56}$Ni mass of $0.14\pm0.02$\,$M_\odot$ and an ejecta mass of $2.2\pm0.6$\,$M_\odot$, while for SN~2016bau, the $^{56}$Ni mass and ejecta mass were $0.055 \pm 0.006\,M_\odot$ and $1.6 \pm 0.3\,M_\odot$, respectively. The photospheric temperature, radius, and velocity evolution for both SNe were also explored. Based on the derived physical quantities from the temporal evolution of these two SNe, we tried to constrain possible powering mechanisms using a semi-analytical model called {\tt MINIM}. We found that the semi-analytical RD model failed to simultaneously fit the early and late phases of both SNe, raising issues related to the validity of such models in the case of stripped-envelope SNe. One solution to such a situation is to fit the early-phase and late-phase data with different sets of model parameters. Another cause for the failure of the RD model can be attributed to the assumption of the constant ejecta density profile. For the MAG model, the fitted parameters seemed to be unphysical in both SNe, especially the very slow initial rotation (for SN~2015ap, $P_{i} \approx 40$\,ms, and for SN~2016bau, $P_{i} \approx 70$\,ms) and the very low initial rotational energy (for both SNe, $E_{p} \approx 10^{49}$\,erg). Also, no signs of these SNe powered by a magnetar mechanism were evident either from photometry or spectroscopy, so this possibility was discarded. 

The spectroscopic behaviour of both SNe was also studied using the present and archival data. These SNe showed unambiguous He~I features from very early to late phases, confirming them to be SNe~Ib. Their spectral features match with those of other well-studied SNe~Ib. Additionally, SN~2015ap closely resembled SN~2008D, which had shown X-ray emission. The spectra of our two SNe at various epochs were modelled using SYN++. For SN~2015ap, the spectral modelling indicated a range of photospheric temperatures and velocities from 13,000\,km\,s$^{-1}$ to 6800\,km\,s$^{-1}$ and from 12,000\,K to 4500\,K (respectively) during the time interval $-7$\,d to +33\,d. For SN~2016bau, these two parameters ranged from 16,000\,km\,s$^{-1}$ to 8000\,km\,s$^{-1}$ and from 9000\,K to 4000\,K (respectively) during the time interval $-14$\,d to +33\,d. The spectra of these two SNe at particular epochs were compared with model spectra of  12, 13, and 17\,$M_{\odot}$ progenitor stars. For SN~2015ap, the 12\,$M_{\odot}$ and 17 $M_{\odot}$ model spectra showed reasonable matches, indicating a progenitor in the mass range 12--17\,$M_{\odot}$; for SN~2016bau, the 12\,M$_{\odot}$ model spectrum could explain the observed spectrum to some extent better than other models, indicating a $\leq 12\,M_{\odot}$ progenitor.

Based on the photometric and spectroscopic properties described above, a 12\,$M_\odot$ ZAMS star was chosen as the possible progenitor for both SNe. This 12\,$M_\odot$ ZAMS progenitor was evolved up to the onset of core collapse using {\tt MESA}. The {\tt MESA} outputs on the onset of core collapse were fed as input to {\tt SNEC} and {\tt STELLA}, which simulate the synthetic explosions. The RD and MAG models were employed using {\tt SNEC} while only the RD model was employed in {\tt STELLA}. Here also, the models failed to fit the late part of the light curve simultaneously. The cause can be attributed to the various assumptions, including the spherically symmetrical explosions and the use of constant ejecta density profiles throughout. Similar to the case of {\tt MINIM}, the MAG model provided unphysical parameters during the {\tt SNEC} analysis.

Our analysis favours a 12\,$M_\odot$ ZAMS star as a possible progenitor for SN~2015ap based on outputs of the models reasonably explaining the bolometric luminosity light curve and the photospheric velocity. However, in case of SN~2016bau, the model velocities are higher by a factor of almost 2, demanding that the ejecta mass be lower by similar factor. This would imply an He star of mass $\sim 2$--2.5\,$M_{\odot}$, which is at the boundary for exploding as an SN. Thus, a slightly lower mass ZAMS star could also be the possible progenitor of SN~2016bau.  Hydrodynamical evolution models of such low-mass stars to reach the stage of core collapse and then undergo synthetic explosions are extremely difficult to perform, but this can be taken as a challenge for the future.   

\section{Conclusions}
\label{sec:Conclusions}
Photometric and spectroscopic analyses of the Lick/KAIT-discovered Type Ib SN~2015ap and another Type Ib SN~2016bau, both having extensive follow-up observations made with various telescopes, are discussed. For both SNe, photometric data corrected for the Milky Way and host-galaxy extinction were used to estimate the quasi-bolometric luminosity light curves and study the photospheric radius, temperature, and velocity evolution. Spectral properties of SN~2015ap and SN~2016bau were then explored in detail. We modelled the spectra, studied the spectral evolution, and compared the spectra at various epochs with those of other well-studied SNe~Ib, which further confirmed that these two SNe are Type Ib SN. 

We attempted to determine the progenitor masses of SN~2015ap and SN~2016bau. Our results support $12\,M_{\odot}$ progenitors for these two SNe. The $12\,M_{\odot}$ ZAMS progenitor was evolved up to the onset of core-collapse using {\tt MESA}. The output of {\tt MESA} was incorporated as input to {\tt SNEC} and {\tt STELLA}, which produced the artificial explosions replicating the actual SN explosions. Considering the decay of Ni and Co to be the most prominent powering mechanism for SNe~Ib, we tried this powering mechanism for SN~2015ap and SN~2016bau, using {\tt SNEC} and {\tt STELLA}. We found that the quasi-bolometric luminosity could nicely be explained by our models, while the velocity evolution obtained from {\tt SNEC} and {\tt STELLA} satisfactorily agrees with the observed one. We also explored the effect of the distribution of nickel mass near the centre and up to near the surface. Lower amounts of nickel were required to match the light curve for the case of centrally distributed nickel in comparison to the case where the nickel was distributed up to near the surface. Based on the above conclusions, our analysis supports a star having $M_{\rm ZAMS} = 12\,M_{\odot}$ as the possible progenitor for SN~2015ap. For SN~2016bau, a slightly lower ZAMS progenitor is expected.

\section*{Acknowledgements}
We are highly thankful to the anonymous referee for providing a very helpful report. We acknowledge Sanyum Channa, Maxime de Kouchkovsky, Andrew Halle, Michael Hyland, Minkyu Kim, Kevin Hayakawa, Kyle McAllister, Jeffrey Molloy, Andrew Rikhter, Benjamin Stahl, and Yinan Zhu  for obtaining some of the Lick  observations. We also thank the {\tt MESA} troubleshooting team, especially Jared Goldberg, for constant guidance. We also acknowledge Van Dyk Schuyler for useful discussion regarding the manuscript. A.A., S.B.P., R.G., and K.M. acknowledge BRICS grant DST/IMRCD/BRICS/Pilotcall/ProFCheap/2017(G). A.A. also acknowledges funds and assistance provided by the Council of Scientific \& Industrial Research (CSIR), India. 
R.O. acknowledges support provided by Japan Society for the Promotion of Science (JSPS) through KAKENHI grant (19J14158). K.M. acknowledges support provided by the Japan Society for the Promotion of Science (JSPS) through KAKENHI grants JP17H02864, JP18H04585, JP18H05223, JP20H00174, and JP20H04737. Support for A.V.F.'s supernova research group has been provided by the TABASGO Foundation, the Christopher R. Redlich Fund, and the U.C. Berkeley Miller Institute for Basic Research in Science (where A.V.F. is a Senior Miller Fellow). Additional support was provide by NASA/{\it HST} grant GO-15166 from the Space Telescope Science Institute (STScI), which is operated by the Associated Universities for Research in Astronomy, Inc. (AURA), under NASA contract NAS 5-26555. 
J.V. is supported by the project ``Transient Astrophysical Objects" (GINOP 2.3.2-15-2016-00033) of the National Research, Development, and Innovation Office (NKFIH), Hungary, funded by the European Union.

Lick/KAIT and its ongoing operation were made possible by donations from Sun Microsystems, Inc., the Hewlett-Packard Company, AutoScope Corporation, Lick Observatory, the U.S. National Science Foundation, the University of California, the Sylvia \& Jim Katzman Foundation, and the TABASGO Foundation. Research at Lick Observatory is partially supported by a generous gift  from Google. Some of the data presented herein were obtained at the W. M. Keck Observatory, which is operated as a scientific partnership among the California Institute of Technology, the University of California, and NASA; the observatory was made possible by the generous financial support of the W. M. Keck Foundation. The Lick and Keck Observatory staff provided excellent assistance with the observations. 

\section*{Data availability}
The photometric and spectroscopic data used in this work can be made available on request to the corresponding author. The {\tt inlist} files to create the {\tt MESA} models and {\tt STELLA} calculations, along with {\tt SNEC} parameters files, can also be made available on request to the corresponding author.







\appendix

\section{Log of spectroscopic and photometric observations}
\begin{table*}
\caption {Spectroscopic observations of SN~2016bau.}
\label{tab:SN2016bau_spec_obs}
\begin{center}
Note: All spectra were taken with the Kast spectrograph on the 3.0\,m Shane telescope at Lick Observatory.
\smallskip
\small\addtolength{\tabcolsep}{-2pt}

\begin{tabular}{c c c c c c}
\hline \hline
UT Date	   &     MJD            \\
           &                    \\
\hline    
2016/03/16  &	57463.719       \\
2016/03/19  &	57466.919       \\
2016/04/03  &	57481.671       \\
2016/04/17  &	57495.902       \\
2016/05/02  &	57510.671       \\
2016/05/16  &	57524.860       \\
2016/05/30  &	57538.741       \\
2016/07/15  &	57584.710       \\
\hline

\end{tabular}
\end{center}
\end{table*}

\begin{table*}
\caption{Photometry of SN~2015ap.}
\centering
\smallskip
\begin{tabular}{c c c c c c c c}
\hline \hline
MJD  	    &  $B$              &  $V$          &  $R$          &  $I$        & Telescope               \\
         &(mag)		    & (mag)             & (mag)         & (mag)         & (mag)                     \\
\hline                           
57274.4258  & 17.35 $\pm$ 0.09 &	17.28 $\pm$  0.07   &   17.07 $\pm$ 0.05  &   16.75 $\pm$  0.06  &    KAIT  \\  
57275.3867  & 16.94 $\pm$ 0.09 &	16.83 $\pm$  0.06   &   16.67 $\pm$ 0.05  &   16.39 $\pm$  0.06  &    KAIT  \\
57276.3516  & 16.40 $\pm$ 0.17 &	16.40 $\pm$  0.11   &   16.39 $\pm$ 0.08  &   16.05 $\pm$  0.07  &    KAIT   \\
57279.4336  & 15.79 $\pm$ 0.11 &	15.84 $\pm$  0.06   &   15.64 $\pm$ 0.05  &   15.39 $\pm$  0.05  &    KAIT   \\
57283.4375  & 15.44 $\pm$ 0.01 &	15.26 $\pm$  0.01   &   15.14 $\pm$ 0.01  &   14.93 $\pm$  0.01  &    Nickel   \\
57285.5078  & 15.64 $\pm$ 0.08 &	15.44 $\pm$  0.04   &   15.20 $\pm$ 0.03  &   14.90 $\pm$  0.03  &    KAIT   \\
57286.4062  & 15.70 $\pm$ 0.04 &	15.48 $\pm$  0.03   &   15.14 $\pm$ 0.02  &   14.87 $\pm$  0.02  &    KAIT   \\ 
57287.4375  & 15.91 $\pm$ 0.05 &	15.47 $\pm$  0.03   &   15.18 $\pm$ 0.03  &   14.87 $\pm$  0.03  &    KAIT   \\
57288.4102  & 15.89 $\pm$ 0.05 &	15.51 $\pm$  0.03   &   15.17 $\pm$ 0.03  &   14.86 $\pm$  0.03  &    KAIT   \\
57290.3555  & 16.07 $\pm$ 0.07 &	15.55 $\pm$  0.05   &   15.20 $\pm$ 0.03  &   14.87 $\pm$  0.03  &    KAIT   \\
57291.3594  & 16.30 $\pm$ 0.09 &	15.58 $\pm$  0.04   &   15.19 $\pm$ 0.03  &   14.88 $\pm$  0.013  &    KAIT   \\
57297.4023  & 17.06 $\pm$ 0.20 &	16.27 $\pm$  0.07   &   15.71 $\pm$ 0.04 &   15.19 $\pm$  0.04  &    KAIT   \\
57298.3906  & 17.30 $\pm$ 0.12 &	16.20 $\pm$  0.04   &   15.75 $\pm$ 0.03  &   15.23 $\pm$  0.03  &    KAIT   \\
57299.5039  & 17.35 $\pm$ 0.15 &	16.53 $\pm$  0.05   &   15.91 $\pm$ 0.03  &   15.22 $\pm$  0.03  &    KAIT   \\
57301.4258  & 17.52 $\pm$ 0.11 &	16.48 $\pm$  0.04   &   15.98 $\pm$ 0.03  &   15.35 $\pm$  0.03  &    KAIT   \\
57302.3125  & 17.43 $\pm$ 0.01 &	16.37 $\pm$  0.01   &   15.86 $\pm$ 0.01  &   15.33 $\pm$  0.01  &    Nickel   \\
57302.3672  & 17.47 $\pm$ 0.08 &	16.52 $\pm$  0.04   &   16.05 $\pm$ 0.03  &   15.46 $\pm$  0.03  &    KAIT   \\
57304.4727  & 17.68 $\pm$ 0.12 &	16.76 $\pm$  0.07   &   16.19 $\pm$ 0.04  &   15.57 $\pm$  0.03  &    KAIT   \\
57305.3828  & 17.90 $\pm$ 0.12 &	16.76 $\pm$  0.05   &   16.27 $\pm$ 0.03  &   15.61 $\pm$  0.03  &    KAIT   \\
57306.3516  & 17.88 $\pm$ 0.18 &	16.84 $\pm$  0.06   &   16.26 $\pm$ 0.04  &   15.65 $\pm$  0.04  &    KAIT   \\
57307.3438  & 17.74 $\pm$ 0.01 &	16.72 $\pm$  0.01   &   16.18 $\pm$ 0.01  &   15.63 $\pm$  0.01  &    Nickel   \\
57307.3945  & 17.82 $\pm$ 0.13 &	16.86 $\pm$  0.04   &   16.27 $\pm$ 0.03  &   15.76 $\pm$  0.04  &    KAIT   \\
57308.3633  & 17.82 $\pm$ 0.12 &	16.83 $\pm$  0.05   &   16.34 $\pm$ 0.03  &   15.75 $\pm$  0.04  &    KAIT   \\
57309.4609  & 17.91 $\pm$ 0.16 &	16.89 $\pm$  0.08   &   16.35 $\pm$ 0.04  &   15.73 $\pm$  0.04  &    KAIT   \\
57317.3047  & 17.98 $\pm$ 0.02 &	17.10 $\pm$  0.01   &   16.64 $\pm$ 0.01  &   15.99 $\pm$  0.01  &    Nickel  \\
57317.4492  & 18.31 $\pm$ 0.24 &	17.25 $\pm$  0.08   &   16.75 $\pm$ 0.05  &   16.02 $\pm$  0.04  &    KAIT   \\
57318.3008  & 18.19 $\pm$ 0.27 &	17.30 $\pm$  0.09   &   16.72 $\pm$ 0.05  &   16.12 $\pm$  0.05  &    KAIT   \\
57319.3984  & 18.22 $\pm$ 0.20 &	17.48 $\pm$  0.12   &   16.80 $\pm$ 0.05  &   16.12 $\pm$  0.07  &    KAIT   \\
57325.4297  & 18.25 $\pm$ 0.38 &	17.14 $\pm$  0.14   &   17.15 $\pm$ 0.09  &   16.19 $\pm$  0.07  &    KAIT   \\
57326.3594  & 18.42 $\pm$ 0.29 &	17.43 $\pm$  0.07   &   17.03 $\pm$ 0.05  &   16.35 $\pm$  0.04  &    KAIT   \\
57327.4297  & 17.98 $\pm$ 0.26 &	17.40 $\pm$  0.12   &            -        &        -             &    KAIT   \\
57330.3906   & 18.37 $\pm$ 0.22 &	17.58 $\pm$  0.08   &   17.20 $\pm$ 0.06  &   16.46 $\pm$  0.05  &    KAIT   \\
57331.3359  & 18.63 $\pm$ 0.22 &	17.72 $\pm$  0.08   &   17.16 $\pm$ 0.04  &   16.59 $\pm$  0.05  &    KAIT   \\
57332.3047  & 18.21 $\pm$ 0.01 &	17.47 $\pm$  0.01   &   17.11 $\pm$ 0.01  &   16.47 $\pm$  0.01  &    Nickel   \\
57332.3555  & 18.31 $\pm$ 0.18 &	17.57 $\pm$  0.07   &   17.21 $\pm$ 0.04  &   16.56 $\pm$  0.04  &    KAIT   \\
57333.3477  & 18.19 $\pm$ 0.17 &	17.73 $\pm$  0.09   &   17.23 $\pm$ 0.06  &   16.56 $\pm$  0.05  &    KAIT  \\
57334.3789  & 18.35 $\pm$ 0.16 &	17.66 $\pm$  0.08   &   17.25 $\pm$ 0.06  &   16.51 $\pm$  0.05  &    KAIT   \\
57337.3281  & 18.33 $\pm$ 0.20 &	18.01 $\pm$  0.18   &   17.33 $\pm$ 0.05  &   16.85 $\pm$  0.06  &    KAIT   \\
57338.2852  & 18.62 $\pm$ 0.28 &	17.79 $\pm$  0.12   &   17.30 $\pm$ 0.05  &   16.73 $\pm$  0.05  &    KAIT   \\
57339.3125  & 18.45 $\pm$ 0.22 &	17.83 $\pm$  0.08   &   17.46 $\pm$ 0.05  &   16.75 $\pm$  0.06  &    KAIT   \\
57339.3438  & 18.28 $\pm$ 0.02 &	17.64 $\pm$  0.01   &   17.27 $\pm$ 0.02  &   16.66 $\pm$  0.01  &    Nickel   \\
57340.3594  & 18.52 $\pm$ 0.18 &	17.76 $\pm$  0.07   &   17.40 $\pm$ 0.06  &   16.73 $\pm$  0.06    &    KAIT   \\
57343.3906  & 18.31 $\pm$ 0.02 &	17.66 $\pm$  0.01   &   17.38 $\pm$ 0.02  &   16.64 $\pm$  0.01  &    Nickel   \\
57347.3164  & 18.51 $\pm$ 0.03 &	17.73 $\pm$  0.01   &   17.51 $\pm$ 0.02  &   17.02 $\pm$  0.02  &    Nickel   \\
57361.2656  & 18.55 $\pm$ 0.02 &	18.14 $\pm$  0.01   &   17.93 $\pm$ 0.02  &   17.19 $\pm$  0.01  &    Nickel   \\
57425.1641  & 19.22 $\pm$ 0.06 &	19.63 $\pm$  0.07   &            -        &        -     &    Nickel   \\
57431.1250  & 19.51 $\pm$ 0.10 &	19.48 $\pm$  0.10   &   19.05 $\pm$ 0.34  &        -     &    Nickel   \\
57444.1328  & 19.67 $\pm$ 0.09 &	20.06 $\pm$  0.10   &   19.82 $\pm$ 0.09  &   18.93 $\pm$  0.06  &    Nickel   \\

\hline                                   
\end{tabular}
\label{tab:optical_observations_SN2015ap}      
\end{table*}

\begin{table*}
\caption {Log of {\it HST} observations of SN~2015ap.\label{tab:HST_log}}
Note: All exposures were 710\,s and there were no detections.
\begin{center}
 
\begin{tabular}{c c c c c c}
\hline \hline
UT Date	   &    Filter        \\
           &                  \\
\hline    
2017/08/16  &    	F555W     \\
2017/02/19  &    	F814W     \\
2016/04/03  &    	F555W     \\
2016/04/17  &    	F814W     \\

\hline
\end{tabular}
\end{center}
\end{table*}

\begin{table*}
\caption{Photometry of SN~2016bau}
\centering
\smallskip
\begin{tabular}{c c c c c c c c}
\hline \hline
MJD  	    &  $B$              &  $V$          &  $R$    & $clear$      &  $I$        & Telescope               \\
         &(mag)		    & (mag)             & (mag)         & (mag)         & (mag)                     \\
\hline                           

57462.384   &    -              &       -           &   -              &  16.896	 $\pm$ 0.190	    &    -            \\
57463.381	& 17.196 $\pm$	 0.160	& 16.691 $\pm$	 0.131	& 16.488 $\pm$	 0.165	& 16.553 $\pm$	 0.226   	& 16.381 $\pm$	 0.124  & KAIT \\
57464.338	& 16.843 $\pm$	 0.119	& 16.408 $\pm$	 0.080	& 16.180 $\pm$	 0.102	& 16.199 $\pm$	 0.103   	& 16.079 $\pm$	 0.086  & KAIT \\
57465.324	& 16.879 $\pm$	 0.120	& 16.192 $\pm$	 0.066	& 16.027 $\pm$	 0.070	& 15.967 $\pm$	 0.131	    & 15.809 $\pm$	 0.106  & KAIT \\
57466.310	& 16.726 $\pm$	 0.134	& 16.005 $\pm$	 0.057	& 15.852 $\pm$	 0.055	& 15.838 $\pm$	 0.105	    & 15.661 $\pm$	 0.063  & KAIT \\
57467.330	& 16.787 $\pm$	 0.253	& 15.890 $\pm$	 0.103	& 15.608 $\pm$	 0.098	& 15.767 $\pm$	 0.156	    & 15.451 $\pm$	 0.128  & KAIT \\
57470.385	& 16.762 $\pm$	 0.165	& 15.741 $\pm$	 0.071	& 15.376 $\pm$	 0.074	& 15.374 $\pm$	 0.061	    & 15.165 $\pm$	 0.091  & KAIT \\
57471.397	& 16.401 $\pm$	 0.176	& 15.589 $\pm$	 0.059	& 15.278 $\pm$	 0.053	& 15.288 $\pm$	 0.062	    & 15.114 $\pm$	 0.056  & KAIT \\
57472.397	& 16.358 $\pm$	 0.140	& 15.477 $\pm$	 0.061	& 15.202 $\pm$	 0.051	& 15.254 $\pm$	 0.065      & 15.031 $\pm$	 0.055  & KAIT \\ 
57473.355	& 16.282 $\pm$	 0.114	& 15.467 $\pm$	 0.049	& 15.196 $\pm$	 0.049	& 15.206 $\pm$	 0.072	    & 14.982 $\pm$	 0.055  & KAIT \\ 
57474.383	& 16.106 $\pm$	 0.315	& 15.464 $\pm$	 0.083	& 15.107 $\pm$	 0.063	& 15.154 $\pm$	 0.075	    & 14.882 $\pm$	 0.114  & KAIT \\ 
57476.373	& 16.079 $\pm$	 0.131	& 15.398 $\pm$	 0.048	& 15.026 $\pm$	 0.049	& 15.067 $\pm$	 0.079	    & 14.806 $\pm$	 0.052  & KAIT \\ 
57477.367	& 16.274 $\pm$ 	 0.088	& 15.348 $\pm$	 0.042	& 15.048 $\pm$	 0.042	& 15.073 $\pm$	 0.046	    & 14.787 $\pm$	 0.042  & KAIT \\ 
57478.332	& 16.340 $\pm$	 0.082	& 15.376 $\pm$	 0.044	& 15.005 $\pm$	 0.042	& 15.055 $\pm$	 0.080	    & 14.744 $\pm$	 0.040  & KAIT \\
57479.395	& 16.290 $\pm$	 0.123	& 15.365 $\pm$	 0.054	& 14.989 $\pm$	 0.053	& 15.039 $\pm$	 0.089	    & 14.750 $\pm$	 0.060  & KAIT \\ 
57480.345	& 16.440 $\pm$	 0.087	& 15.422 $\pm$	 0.034	& 15.037 $\pm$	 0.038	& 15.062 $\pm$	 0.045	    & 14.753 $\pm$	 0.056  & KAIT \\
57481.338	& 16.487 $\pm$	 0.207	& 15.444 $\pm$	 0.070	& 15.102 $\pm$	 0.065	& 15.105 $\pm$	 0.073	    & 14.713 $\pm$	 0.062  & KAIT \\ 
57482.341	& 16.801 $\pm$	 0.142  & 15.527 $\pm$	 0.053	& 15.057 $\pm$	 0.045	& 15.132 $\pm$	 0.065	    & 14.784 $\pm$	 0.049  & KAIT \\
57485.358	& 17.453 $\pm$	 0.207	& 15.691 $\pm$	 0.115	& 15.237 $\pm$	 0.093	& 15.320 $\pm$	 0.090	    & 15.039 $\pm$	 0.090  & KAIT \\
57491.352	& 17.914 $\pm$	 0.243	& 16.282 $\pm$	 0.118	& 15.730 $\pm$	 0.119	& 15.744 $\pm$	 0.141	    & 15.224 $\pm$	 0.116  & KAIT \\
57492.352	& 17.765 $\pm$	 0.159	& 16.376 $\pm$	 0.103	& 15.776 $\pm$	 0.085	& 15.824 $\pm$	 0.137	    & 15.262 $\pm$	 0.073  & KAIT \\
57494.257	& 17.979 $\pm$	 0.169	& 16.484 $\pm$	 0.108	& 15.896 $\pm$	 0.111	& 15.941 $\pm$	 0.148	    & 15.392 $\pm$	 0.125  & KAIT \\
57495.299	& 18.142 $\pm$	 0.134	& 16.553 $\pm$ 	 0.069	& 15.914 $\pm$	 0.071	& 16.018 $\pm$	 0.095	    & 15.427 $\pm$	 0.064  & KAIT \\
57496.272   & 18.106 $\pm$	 0.131	& 16.603 $\pm$	 0.065	& 15.970 $\pm$	 0.063	& 16.048 $\pm$	 0.093	    & 15.445 $\pm$	 0.058  & KAIT \\
57497.268	& 18.194 $\pm$	 0.186	& 16.721 $\pm$	 0.086	& 15.966 $\pm$	 0.065	& 16.100 $\pm$	 0.084	    & 15.492 $\pm$	 0.067  & KAIT \\
57498.330   & 18.137 $\pm$	 0.145	& 16.646 $\pm$	 0.094	& 16.065 $\pm$	 0.087	& 16.129 $\pm$	 0.114	    & 15.557 $\pm$   0.079  & KAIT \\
57499.317	& 18.147 $\pm$	 0.329	& 16.675 $\pm$	 0.149	& 16.111 $\pm$	 0.141	& 16.193 $\pm$	 0.230	    & 15.555 $\pm$	 0.131  & KAIT \\
57502.338	& 18.204 $\pm$	 0.277	& 16.935 $\pm$	 0.170	& 16.180 $\pm$	 0.204	& 16.201 $\pm$	 0.147	    & 15.674 $\pm$	 0.095  & KAIT \\
57504.271	& 18.340 $\pm$	 0.173	& 16.893 $\pm$	 0.111	& 16.291 $\pm$	 0.114	& 16.313 $\pm$	 0.165	    & 15.738 $\pm$ 0.095  & KAIT \\
57505.327	&     -             &         -         &         -         & 16.297	  $\pm$ 0.148	    &     -           & KAIT \\
57507.265	& 18.362 $\pm$	 0.179	& 16.974 $\pm$	 0.109	& 16.372 $\pm$	 0.116	& 16.408 $\pm$	 0.184	    & 15.803 $\pm$	 0.089  & KAIT \\
57508.241	& 18.348 $\pm$	 0.248	& 16.993 $\pm$	 0.197	& 16.445 $\pm$	 0.210	& 16.408 $\pm$	 0.237	    & 15.837 $\pm$	 0.163  & KAIT \\
57509.285	& 18.414 $\pm$	 0.156	& 17.023 $\pm$	 0.110	& 16.401 $\pm$	 0.119	& 16.441 $\pm$	 0.181	    & 15.825 $\pm$	 0.098  & KAIT \\
57510.229	&	-               &      -            &	      -         & 16.458	 $\pm$ 0.172	    &     -             & KAIT \\
57511.236	& 18.498 $\pm$	 0.154	& 17.092 $\pm$	 0.121	& 16.484 $\pm$	 0.086	& 16.447 $\pm$	 0.178	    & 15.848 $\pm$	 0.087  & KAIT \\
57513.195	& 18.490 $\pm$	 0.417	& 17.150 $\pm$	 0.163	& 16.552 $\pm$	 0.134	& 16.517 $\pm$	 0.195	    & 15.929 $\pm$	 0.112  & KAIT \\
57517.272	& 18.501 $\pm$	 0.189	& 17.238 $\pm$	 0.104	& 16.619 $\pm$	 0.116	& 16.594 $\pm$	 0.221	    & 16.001 $\pm$	 0.096  & KAIT \\
57518.210	& 18.590 $\pm$	 0.151	& 17.221 $\pm$	 0.129	& 16.668 $\pm$	 0.114	& 16.591 $\pm$	 0.139	    & 15.993 $\pm$	 0.094  & KAIT \\
57519.223	& 18.514 $\pm$	 0.156	& 17.211 $\pm$	 0.094	& 16.671 $\pm$	 0.095	& 16.612 $\pm$ 	 0.127	    & 16.043 $\pm$	 0.101  & KAIT \\
57520.193	& 18.453 $\pm$	 0.167	& 17.284 $\pm$	 0.148	& 16.669 $\pm$	 0.099	& 16.651 $\pm$	 0.349	    & 16.039 $\pm$	 0.102  & KAIT \\
57521.203	& 18.651 $\pm$	 0.179	& 17.276 $\pm$	 0.126	& 16.737 $\pm$	 0.121	& 16.699 $\pm$	 0.119	    & 16.057 $\pm$	 0.093  & KAIT \\
57522.186	& 18.638 $\pm$	 0.215	& 17.359 $\pm$	 0.186	& 16.744 $\pm$	 0.190	& 16.727 $\pm$	 0.263	    & 16.074 $\pm$	 0.155  & KAIT \\
57523.259	& 18.510 $\pm$	 0.235	& 17.396 $\pm$	 0.194	& 16.697 $\pm$	 0.149	& 16.672 $\pm$	 0.322	    & 16.029 $\pm$	 0.118  & KAIT \\
57524.206	& 18.643 $\pm$	 0.147	& 17.346 $\pm$	 0.121	& 16.717 $\pm$	 0.116	& 16.753 $\pm$	 0.229	    & 16.068 $\pm$	 0.102  & KAIT \\
57525.193	& 18.675 $\pm$	 0.168	& 17.392 $\pm$	 0.090	& 16.802 $\pm$	 0.118	& 16.729 $\pm$	 0.134	    & 16.117 $\pm$	 0.110  & KAIT \\
57526.218	& 18.464 $\pm$	 0.205	& 17.358 $\pm$	 0.120	& 16.748 $\pm$	 0.113	& 16.739 $\pm$	 0.243	    & 16.097 $\pm$	 0.101  & KAIT \\
57527.200	& 18.490 $\pm$	 0.166	& 17.401 $\pm$ 	 0.092	& 16.824 $\pm$	 0.116	& 16.846 $\pm$	 0.201	    & 16.115 $\pm$  	 0.089  & KAIT \\
57531.212	& 18.761 $\pm$	 0.175	& 17.450 $\pm$	 0.096	& 16.854 $\pm$	 0.102	& 16.832 $\pm$	 0.221	    & 16.214 $\pm$	 0.114  & KAIT \\
57533.227	& 18.663 $\pm$	 0.143	& 17.541 $\pm$ 	 0.109	& 16.914 $\pm$	 0.143	& 16.848 $\pm$	 0.249	    & 16.209 $\pm$	 0.075  & KAIT \\
57534.202	& 18.583 $\pm$	 0.130	& 17.519 $\pm$	 0.085	& 16.984 $\pm$	 0.133	& 16.933 $\pm$	 0.264  	& 16.271 $\pm$	 0.102  & KAIT \\
57535.215	& 18.683 $\pm$	 0.163	& 17.558 $\pm$	 0.110	& 16.998 $\pm$	 0.112	& 16.975 $\pm$	 0.277	    & 16.269 $\pm$	 0.105  & KAIT \\
57536.219	& 18.693 $\pm$	 0.177	& 17.587 $\pm$	 0.101	& 16.981 $\pm$	 0.103	& 16.958 $\pm$	 0.257	    & 16.275 $\pm$	 0.076  & KAIT \\
57537.232	& 18.818 $\pm$	 0.224	& 17.562 $\pm$	 0.098	& 17.001 $\pm$	 0.093	& 17.010 $\pm$	 0.270	    & 16.257 $\pm$	 0.093  & KAIT \\
57541.231	& 18.793 $\pm$	 0.151	& 17.682 $\pm$	 0.086	& 17.137 $\pm$	 0.117	& 16.981 $\pm$	 0.263	    & 16.355 $\pm$	 0.097  & KAIT \\
57552.211	& 18.943 $\pm$	 0.150	& 17.819 $\pm$	 0.124	& 17.316 $\pm$	 0.100	& 17.231 $\pm$	 0.070	    & 16.525 $\pm$	 0.088  & KAIT \\
57572.191	& 19.171 $\pm$	 0.331	& 18.126 $\pm$ 	 0.132	& 17.587 $\pm$	 $\pm$0.117	& 17.422 $\pm$	 0.139	& 16.842 $\pm$	 0.118  & KAIT \\
57574.190	&      -                &      -                &         -                      & 17.403 $\pm$   0.091	&      -                & KAIT \\
57576.191	&      -                &      -                &         -                      & 17.544 $\pm$ 0.519	&      -                & KAIT \\
57471.326	& 16.267 $\pm$	 0.022	& 15.589 $\pm$	 0.015	& 15.274$\pm$       0.019    	& 	-	                & 15.065 $\pm$	 0.027  & KAIT \\
57484.382	& 16.823 $\pm$	 0.029	& 15.501 $\pm$	 0.034	& 15.213 $\pm$	 0.014    	& 	-	                & 14.842 $\pm$	 0.022  & KAIT \\
57498.264	& 17.822 $\pm$	 0.081	& 16.741 $\pm$	 0.057  & 16.069 $\pm$	 0.031	    &   -	                & 15.551 $\pm$	 0.035  & KAIT \\
57511.285	& 18.314 $\pm$	 0.376	& 17.054 $\pm$	 0.046	& 16.397 $\pm$	 0.025   	&   -	                & 15.843 $\pm$	 0.616  & KAIT \\
57535.248	& 18.563 $\pm$	 0.058	& 17.544 $\pm$	 0.035  & 16.948 $\pm$	 0.044    	& 	-	                & 16.289 $\pm$	 0.072  & KAIT \\
57541.241	& 18.555 $\pm$	 0.719	& 17.565 $\pm$	 0.070	& 17.014 $\pm$    0.034    	&   -	                & 16.347 $\pm$	 0.028  & KAIT \\
57546.216	& 18.816 $\pm$	 0.229	& 17.691 $\pm$	 0.070	& 17.139 $\pm$	 0.028    	&   -	                & 16.447 $\pm$	 0.099  & KAIT \\
57587.185	& 19.431 $\pm$	 0.701	& 18.453 $\pm$	 0.309	&      -                         &   -                   &       -               & KAIT      \\
\hline                                   
\end{tabular}
\label{tab:optical_observations_2016bau}      
\end{table*}



\bsp	
\label{lastpage}
\end{document}